\documentclass[sigconf]{acmart}

\usepackage{xcolor}

\usepackage{textcomp}

\usepackage{tabularx}
\usepackage{graphicx}
\usepackage{multirow}

\usepackage[shortlabels]{enumitem}

\usepackage{algorithm}
\usepackage{algorithmic}
\usepackage[tight,footnotesize]{subfigure}
\usepackage{caption}

\usepackage{fixltx2e}

\usepackage{stfloats}

\usepackage{url}

\usepackage{soul}

\copyrightyear{2022}
\acmYear{2022}
\setcopyright{rightsretained}
\acmConference[ACSAC '22]{Annual Computer Security Applications Conference}{December 5--9, 2022}{Austin, TX, USA}
\acmBooktitle{Annual Computer Security Applications Conference (ACSAC '22), December 5--9, 2022, Austin, TX, USA}
\acmDOI{10.1145/3564625.3567986}
\acmISBN{978-1-4503-9759-9/22/12}

\begin{document}

%
\pagestyle{plain}


\title{NeuGuard: Lightweight \underline{Neu}ron-\underline{G}uided \underline{D}efense against Membership Inference Attacks}

\author{Nuo Xu}
\affiliation{%
  \institution{Lehigh University}
    \city{Bethlehem}
  \state{PA}
  \country{USA}
}
\email{nux219@lehigh.com}

\author{Binghui Wang}
\affiliation{%
  \institution{Illinois Institute of Technology}
      \city{Chicago}
  \state{IL}
  \country{USA}
}
\email{bwang70@iit.edu}

\author{Ran Ran}
\affiliation{%
  \institution{Lehigh University}
    \city{Bethlehem}
  \state{PA}
  \country{USA}
}
\email{rar418@lehigh.com}
\author{Wujie Wen}
\affiliation{%
  \institution{Lehigh University}
    \city{Bethlehem}
  \state{PA}
  \country{USA}
}
\email{wuw219@lehigh.com}

\author{Parv Venkitasubramaniam}
\affiliation{%
  \institution{Lehigh University}
  \city{Bethlehem}
  \state{PA}
  \country{USA}
}
\email{pav309@lehigh.com}

\begin{abstract}
Membership inference attacks (MIAs) against machine learning models lead to serious privacy risks for the training dataset used in the model training. The state-of-the-art defenses against MIAs often suffer from poor privacy-utility balance and defense generality, as well as high training or inference overhead. To overcome these limitations, in this paper, we propose a novel, lightweight and effective \underline{Neu}ron-\underline{G}uided \underline{D}efense method named \textit{NeuGuard} against MIAs. Unlike existing solutions which either regularize all model parameters  in training or noise model output per input in real-time inference, \textit{NeuGuard} aims to wisely guide the model output of training set and testing set to have close distributions through a fine-grained neuron regularization. That is, restricting the activation of output neurons and inner neurons in each layer simultaneously by using our developed class-wise variance minimization and layer-wise balanced output control.  
We evaluate \textit{NeuGuard} and compare it with state-of-the-art defenses against two neural network based MIAs, five strongest metric based MIAs including the newly proposed label-only MIA on three benchmark datasets. 
Extensive experimental results show that \textit{NeuGuard} outperforms the state-of-the-art defenses by offering much improved utility-privacy trade-off,  generality, and overhead. 
Our code is publicly available at \url{https://github.com/nux219/NeuGuard}.
\end{abstract}


\maketitle

%


\section{Introduction}
Machine learning (ML) has achieved extraordinary success in many fields, spanning daily image classification, object detection, and privacy- and security- sensitive medical diagnosis~\cite{ozbulak2019impact}, biometric authentication~\cite{ndia2022}, and autonomous vehicles~\cite{versprille2015researchers}.
Such success mainly depends on training ML models with the large-scale domain-specific datasets that may contain crucial confidential and private information, e.g. personal medical records, human faces. Unfortunately, many recent studies have revealed that such data information can be retrieved from the trained ML models by various attacks, such as attribute inference attack, model inversion attack, and membership inference attack~\cite{jia2017attriinfer,zhang2016understanding,fredrikson2015model,fredrikson2014privacy,tramer2016stealing,shokri2017membership,
melis2019exploiting,truex2019demystifying, choquette2021label, li2021label,truex2018towards}.  

Among these attacks, membership inference attack (MIA)~\cite{shokri2017membership} has been attracting ever increasing attention, 
with the goal of identifying whether a given data sample is used for training the target ML model or not. For instance, if a model is trained using a specialized medical database, e.g. data pertaining to individuals' disease, a successful MIA could reveal the identity of a person having the disease. 
In essence, MIAs are built upon the fact that a model's response to the member (training set) and non-member (testing set) can be different by nature. This difference allows the attacker to either train neural network (NN) based binary classifiers~\cite{shokri2017membership,nasr2018machine} or use non-NN metric based approaches~\cite{salem2018ml,song2021systematic,choquette2021label} for accurate member inferring. Given the difficulty of fundamentally eliminating such a difference, defending against MIAs can be challenging. 

There exist many 
studies to address MIAs, including provable defense like differential privacy (DP)~\cite{abadi2016deep,rahman2018membership}, and non-provable defense like training time defense with dropout ~\cite{salem2018ml}, $L_2$ norm ~\cite{shokri2017membership}, MIA-dedicated min-max adversary regularization~\cite{nasr2018machine}, distillation based defense~\cite{shejwalkar2021membership} and inference time defense via output perturbation~\cite{jia2019memguard}. Compared with provable defense like DP that is known for provable privacy but extremely low model utility~\cite{shejwalkar2021membership,abadi2016deep,jayaraman2019evaluating} (see Section~\ref{related_work}), the non-provable defenses offer privacy empirically but much better utility. However, as we shall show in Section~\ref{mot}, these solutions are still far from satisfactory in terms of model utility-defense effectiveness balance, defense generality against a wide range of MIAs, and training and inference overhead. \textit{The underlying reasons are that existing solutions either regularize and control all model parameters in a coarse-grained manner with no guarantee of minimizing the model output distribution difference between training set and testing set, or directly manipulate output scores for achieving similar distributions by expensively searching and adding sample-specific perturbation noise during the real-time inference.} To this end, there has been lack of systematic studies on an important facet, that is, \ul{effectively guiding the model to produce output that would ensure that the output distribution for training set and test set are very close, with much better defense effectiveness and generality at marginal utility loss and overhead.}   

To achieve this goal, in this paper, we propose to create a new paradigm of safeguarding MIAs from a radically different perspective. Our basic idea is to \textbf{develop lightweight and fine-grained neuron-level regularization to simultaneously guide and orchestrate the final output neurons and hidden neurons (or intermediate features) for producing an output confidence score distribution indistinguishable between the training set and testing set.}
\textit{\ul{The key rationale is that}}: if we can deliberately and largely reduce the space of distributions obtained over confidence score for training data by 
developing privacy-dedicated neuron-level training regularization, 
then such trained model will naturally confine output score distribution to a very limited space regardless of input (either training or testing data). In this way, the output score distribution of testing data produced by this model is also restricted to the similar space with a much smaller variance (as confirmed in Table~\ref{mean_var_cifar100} and ~\ref{tab:kld_all}), and thus is close to that of training data. While there still exist some score distribution differences between the training and testing data, the gap between them can be much smaller compared with existing solutions, thereby offering much better defense effectiveness against a variety of MIAs.
However, the challenges lie in designing a learnable target output distribution with much reduced space and neuron regularizations workable for output and inner neurons with minimized utility loss and overhead. 

To overcome these challenges, we design a class-wise variance minimization regularization that directly acts on the final output neurons to minimize the variance of output score in training, along with layer-wise balanced output control regularization for inner neurons to further guide the learning of intermediate features.
As a result, the output confidence score of any input can be more evenly distributed with all label's confidence values close to the mean value while still maintaining the prediction correctness (see Fig.~\ref{fig_flow}(a)). In this way, we successfully restrict the value range and distribution 
of the output score, and hence reduce the
model response difference between the training set and testing set. We name such a \ul{Neu}ron-\ul{G}uided \ul{D}efense against MIAs as \textbf{\textit{NeuGuard}} in this work.

We evaluate our proposed \textit{NeuGuard} against both NN based attacks and the latest metric based attacks (including the  label-only attack), and compare it with state-of-the-art defenses~\cite{nasr2018machine,jia2019memguard,song2021systematic} on three datasets. Experimental results show \textit{NeuGuard} delivers by far the best utility-defense trade-off among the known defense mechanisms.
Our \textit{NeuGuard} achieves lower attack accuracy against all attacks 
at marginal utility drop and much lower training and inference overhead. Moreover, \textit{NeuGuard} also provides a much better generality comparing with prior solutions, that is, models protected by \textit{NeuGuard} are resilient to all kinds of NN and Non-NN metric based MIAs.  
Our contributions are summarized as follows:
\begin{itemize}[leftmargin=*]
 \item We, for the first time, investigate the difference between neural network based membership inference attacks that use sorted output confidence scores or unsorted scores with label information as two different attack inputs,
 and find that existing defenses workable for one attack often do not work for the other. 
\item We develop a novel, simple and effective neuron-guided defense--\textit{NeuGuard}, to explicitly reduce the model output distribution difference between training set and testing set. It promises the best privacy and utility trade-off.   
\item We extensively evaluate our defense against NN based membership inference attacks and metric based attacks, outperforming existing defenses on three real-world datasets. 
 \item We explore why defenses cannot reduce the metric-based attacks to random guessing, and perform analytical and experimental studies on the upper bound of defense effectiveness.
\end{itemize}

\section{Background}
\label{background}

\subsection{Membership inference attacks}
\label{intro_MIA}
The MIA
attempts to infer whether a given data is from the training set or not that is used to train a target model ~\cite{shokri2017membership,song2021systematic,song2019privacy,yeom2018privacy,salem2018ml,choquette2021label}. 
There exist many MIAs with different evaluation measurements and attack capabilities. In this work, we discuss two major classes of MIAs: 
\emph{neural network based attacks} and \emph{metric based attacks}.


\subsubsection{Neural network based attacks}
\label{intro_nn_MIA}
They aim to identify the given data's membership using neural networks and are performed in the white-box or black-box fashion.  
White-box MIAs assume the attacker has full access to the target model~\cite{nasr2019comprehensive,leino2020stolen}, e.g. model architecture and parameters. 
Black-box MIAs assume the attacker can only observe the output confidence scores of the target model,
which are more realistic and the focus of this work. 

Given a data sample and its confidence scores outputted by a target model, neural network based attack trains 
a binary classifier 
to determine whether the data belongs to the training set or not.
In particular, there are two kinds of black-box attacks, and they major differ in 
whether 
the output confidence scores are sorted 
before the attack or not. 
\textit{We would like to emphasize that the existing research does not explicitly differentiate these two kinds of attacks, and we notice that  
a defense being effective to one attack is often not effective to the other type
(Please see Section~\ref{nn_based_attack_evaluation})}. We, for the first time, notice the difference between these two attacks, and our proposed defense \textit{NeuGuard} shows better defense generality against both the attacks.

\textbf{NN based attack with sorted input}: 
The first NN based attack takes the sorted output confidence scores as the attack input. We name this attack \textit{sorted NN attack} in short. 
The binary attack classifier focuses on learning the difference between training set and testing set while ignoring the correctness of the class prediction.
Training sorted NN attack can be realized by different approaches: 1) using a shadow model to represent the target model and training it by performing MIA with multiple class-wise attack classifiers~\cite{shokri2017membership}; 2) using a general attack classifier for all classes and achieve similar attack accuracy ~\cite{salem2018ml}. 

\textbf{NN based attack with unsorted input}: 
The second NN based attack uses both the unsorted output confidence scores and label information as attack inputs~\cite{nasr2018machine}. We call it \textit{unsorted NSH attack} for short in this paper. 
This attack uses three neural network models to construct the binary classifier. One model is used to receive the unsorted confidence scores, and one takes the one-hot encoded class label as input. The last one concatenates the outputs from the other two models and generates a single probability, to determine whether the given data is a member of the training set or not. 

\subsubsection{Metric based attacks}
\label{metricbased_attack}
Unlike neural network based attacks, metric-based attacks directly 
compute customized metrics using prediction confidence scores or only the label information to infer membership or non-membership. 
We introduce the four state-of-the-art (strongest) metric based attacks and the newly proposed label only attack for our later evaluation~\cite{song2021systematic,choquette2021label}.  
Let $(x,y)$ be a data sample with features $x$ and label $y$ and $F$ be a NN model.

\noindent \textbf{Prediction correctness based MIA.} This attack infers the membership based on whether the given input data is correctly classified by the target model or not~\cite{yeom2018privacy}.
The associated metric $\mathcal{M}_{\text {corr }}$ is defined as:
\begin{equation}\label{}
\mathcal{M}_{\text {corr }}(F;x)=I(\operatorname{argmax} F(x)=y),
\end{equation}%
where $I(\cdot)$ is an indicator function.

\noindent \textbf{Prediction confidence based MIA.}
This attack determines whether an input data 
is from the training set or not by 
comparing the most significant confidence score with a preset threshold. 
The attack is first designed by ~\cite{salem2018ml} using a single threshold for all classes and ~\cite{song2021systematic} improves it by applying class-wise thresholds to reduce the impact of confidence differences among classes. 
The associated metric  $\mathcal{M}_{\text {conf}}$ is defined as:
\begin{equation}
\label{eqn:pc}
\mathcal{M}_{\text {conf}}(F;x)=I( F(x)_{y} \geq \tau_{y}),
\end{equation}%
where $\tau_{y}$ represents the threshold for the class $y$.

\noindent \textbf{Prediction entropy based MIA.}
The entropy based MIA attack ~\cite{song2021systematic} 
is based on the fact that the testing set prediction entropy should be much larger than the training set.  
It identifies the input data $x$ as a member if the prediction entropy is smaller than a preset threshold (e.g., $\hat{\tau}_{y}$). The associated metric  $\mathcal{M}_{\text {entr}}(F;x)$ is defined as: 
\begin{equation}\label{}
\mathcal{M}_{\text {entr}}(F;x)=I(-\textstyle\sum_{i=0}^{k}  F(x)_{i} \log \left(F(x)_{i}\right) \leq \hat{\tau}_{y}).
\end{equation}

\noindent \textbf{Modified prediction entropy based MIA.}
Prediction entropy attack has a major limitation in that it doesn't contain any label information~\cite{song2021systematic}. As a result, 
both correct and incorrect predication with high score can lead to zero entropy values.
Modified prediction entropy~\cite{song2021systematic} fixes this issue that only high probability correct prediction can lead modified entropy to 0.
Then such modified entropy $\operatorname{ME}(F(x), y)$ is presented as: $\operatorname{ME}(F(x), y)=$ $-\left(1-F(x)_{y}\right) \log \left(F(x)_{y}\right) -{\textstyle\sum_{i \neq y}} F(x)_{i} \log \left(1-F(x)_{i}\right) $.
The adversary determines an input data as a member if 
the above modified value is smaller than the preset class-related threshold--$\check{\tau}_{y}$ for class $y$. The associated metric  $\mathcal{M}_{\text {Mentr}}(F;x)$ is defined as: 
\begin{equation}\label{eqn:mpe}
\mathcal{M}_{\text {Mentr }}(F;x)=I(\operatorname{ME}(F(x), y) \leq \check{\tau}_{y})
\end{equation}
\textbf{Label-only MIA~\cite{choquette2021label}.} It is a newly proposed MIA that tries to infer membership or non-membership with only the label information. 
Specifically, for a data example $(x,y)$, it first generates an adversarial example $x'$ (an example close to $x$ but has a wrong predicted label) 
and calculates the $l_2$-distance to a
model’s decision boundary as $dist_F(x', y)$. Then $x$ is predicted as a member if $dist_F(x', y) > \tau$ ($\tau$ is a threshold) since a training data should exhibit higher robustness than testing data.
In our evaluation, we adopt the strongest C$\&$W~\cite{carlini2017towards} based label-only MIA to generate adversarial examples. 
This attack serves as the upper bound of the label-only MIAs, as it uses the actual model's information (e.g. gradient) to generate ideal adversarial examples for MIAs (white-box setting)~\cite{choquette2021label}.

\subsection{The state-of-the-art defense methods}
\label{SOTA-de}
We mainly introduce two kinds of most representative and powerful defenses for detailed experimental comparisons in Section~\ref{nn_based_attack_evaluation} and \ref{metric_base_evaluation}: training regularization and inference output perturbation.

\subsubsection{Training regularization defense}
Many works ~\cite{chen2020gan,leino2020stolen,salem2018ml,shokri2017membership,yeom2018privacy} have pointed out that model overfitting is the main reason that makes MIAs effective. Based on this observation, adding 
a regularizer during training to reduce  overfitting can be an effective way to defend against MIA.
The common regularization methods include L2-norm
regularization~\cite{shokri2017membership}, dropout ~\cite{srivastava2014dropout}, model stacking~\cite{salem2018ml}, early stop~\cite{song2021systematic}.
However, they mainly focus on model utility, and thus are unable to greatly reduce membership inference attack accuracy.

Besides, adversary regularization~\cite{nasr2018machine} (\textbf{AdvReg} in short) is specifically designed to defense against MIAs. The method introduces an NN based membership inference classifier during training to achieve the defense.
The training process needs to simultaneously minimize the target model's loss and the attack classifier's accuracy over the training set $D_{\mathrm{tr}}$: 
\begin{equation}\label{}
\small
\underset{\theta}{\operatorname{argmin}} \frac{1}{\left|D_{\mathrm{tr}}\right|} \textstyle\sum_{\mathbf{x} \in D_{\mathrm{tr}}} L(F(x), y)+\lambda \log (I(F(x), y)),
\end{equation}
where $\lambda$ is a hyperparameter to balance the privacy risk and original classification task, and $I(\cdot)$ is the attack classifier.

\subsubsection{Inference output perturbation defense}
Another effective way is to directly modify the output confidence scores such that the information is hidden from the adversary. 
The basic idea of the best existing defense method \textbf{MemGuard} ~\cite{jia2019memguard} 
is to add carefully calculated noise into the output confidence scores and turn the output into adversarial examples~\cite{athalye2018obfuscated} to mislead the attack classifier for each input. Since the noise is crafted and added at the inference stage, this method does not influence the training process and maintains the target model's accuracy.

\section{Motivation and threat model}
\label{motivation}
\subsection{Motivation}
\label{mot}
We aim to satisfy the following requirements to achieve a good defense against MIAs. Existing solutions, however, are incapable of addressing challenges to meet such requirements: 
\begin{itemize}[leftmargin=*]
  \item \textbf{Defense effectiveness:} A good defense 
  shall reduce MIA attack accuracy as close as to a random guess, i.e., 50\%.

  \item \textbf{Defense generalizability:} A good defense shall be able to defend against different types of attacks considering the uncertainty of experiencing which MIA in practice. 
  \item \textbf{Utility loss:} A good defense shall maintain the target model's accuracy on unseen data as much as possible.  
  A defense causing a large accuracy drop is not desirable.

  \item \textbf{Overhead:} A good defense shall be lightweight and not incur significant overhead in training or inference.  
\end{itemize}

\newcommand{\hollowstar}{\text{\ding{73}}}

Table~\ref{defense_requirement} briefly evaluates the state-of-the-art defense methods discussed in Section~\ref{SOTA-de} using the above four criteria.
These results are summarized based on previous works ~\cite{nasr2018machine,jia2019memguard,song2021systematic} and our experiment results (details in Section \ref{comb_res} and \ref{metric_base_evaluation}).
According to the table, existing defenses all have limitations in some aspects.

{\bf Dropout} reduces model overfitting, but exhibits limited MIA defense, e.g. only slightly better than the baseline.  
{\bf Early stopping} sacrifices model utility to improve defense efficiency, but obtains a sub-optimal utility-defense trade-off.

{\bf AdvReg} leverages a min-max game theoretic method to train the model against MIAs. However, it cannot provide an effective defense while maintaining the utility in practice~\cite{shejwalkar2021membership,song2021systematic}. 
The defense takes effects on the Texas100 when $\lambda \ge 3$, but it can lead to 8\% - 18\% accuracy drop (see Table~\ref{advreg_texas100} in Appendix). Furthermore, ~\cite{song2021systematic} evaluates AdvReg in metric based MIAs and shows it is no better than  
early stopping under a similar test accuracy.
The reason is that AdvReg directly incorporates the MIA classifier as a regularization term into the training but it may not have an explicit goal to regularize the model towards lowering MIAs. Besides, the difficulty of jointly optimizing the MIA classifier and original task also leads to considerable training overhead comparing to regular training.

\begin{table}[t]
\renewcommand{\arraystretch}{}
\caption{\small Comparing existing MIA defenses.}
\label{defense_requirement}
\vspace{-10pt}
\centering
\resizebox{\linewidth}{!}{
\begin{tabular}{l|llll}
\hline
Method & \begin{tabular}[l]{@{}l@{}}Defense\\ effectiveness\end{tabular} & \begin{tabular}[l]{@{}l@{}}Defense\\ Generalizability\end{tabular} & \begin{tabular}[l]{@{}l@{}}Model\\ utility\end{tabular} & Overhead \\ \hline
Normal training & $--$   & $--$  & $++$ & No \\
Dropout ~\cite{salem2018ml} & $-$  & $-$ & $++$ & Low  \\
Early stopping ~\cite{song2021systematic} & $+$  &  $+$ & $-$  & No \\
AdvReg ~\cite{nasr2018machine}     &  $+$  &  $+$ & $-$  & Medium (Training) \\
MemGuard ~\cite{jia2019memguard}   &  $++$ &  $-$ & $++$ & High (Inference)\\ \hline
\end{tabular}
}
{ `$++$' indicates the best, `$--$' means the worst. \par}
\vspace{-15pt}
\end{table}

{\bf MemGuard} performs defense at the inference stage such that the model utility is not affected and there is no training overhead. However, it suffers from the following limitations which hinder it from satisfying the given defense requirements: 
\textbf{First}, MemGuard causes huge inference overhead as it needs to run calculations for many times (e.g. up to 300$\times$) for each input sample to find out the best noise output to defend against the attack.
\textbf{Second}, its defense performance is highly dependent on the given trained model.
It may fail to find the noisy outputs for the target model to effectively defend against the attack, leaving a high attack accuracy.   
\textbf{Third}, MemGuard cannot provide general protection against different attacks. If it uses a sorted NN classifier as the defense classifier, it cannot defend the attack with the unsorted NSH model and vice versa (details in Section~\ref{nn_based_attack_evaluation}). ~\cite{song2021systematic} further evaluates MemGuard under metric based attacks and shows that the defense works to a limited degree and is not much better than the baseline model.
In addition, as demonstrated in ~\cite{choquette2021label}, post-processing output scores like MemGuard fails to defend against label-only attacks as model produced raw scores remain unchanged (i.e., unchanged distance between input to decision boundary to infer members in label-only).

Based on the above limitation analysis of existing defenses, we believe a better way to defend against MIAs is to develop a method that is unlike AdvReg and has fine-grained regularization control guided by more specific objectives with the guarantee of lowered MIAs. The defense method should also maintain the model utility and provide a good utility-defense trade-off, work under different MIAs (e.g., sorted NN, unsorted NSH, metric-based attacks), and has much lower overhead than existing defenses (e.g. MemGuard).

\vspace{-5pt}
\subsection{Threat model}
\label{threatmodel}
In this work, we adopt a threat model consistent with previous defenses~\cite{nasr2018machine, jia2019memguard}. We assume model providers have a private training dataset (e.g., financial records, healthcare dataset, or location dataset). 
They train a machine learning model with the private dataset and deploy the trained model as a service provided by the cloud server such as Machine-Learning-As-A-Service (MLaaS). Users can perform inferences and receive prediction results (score or label) from the server.
We also assume the model providers can apply defense  methods in the training and/or inference stages. The providers' goal is to make the model have satisfactory test performance and capable of defending against various membership inference attacks. 
The attackers aim to infer the membership of the private data in the training set from the deployed model. 
We assume the attacker could receive the output confidence scores or predicted labels directly generated by the target model from the server, instead of the shadow models estimated by attackers locally, to maximize MIA effectiveness for attackers. However, attackers cannot access the model itself or the model parameters. 

\textbf{Partial knowledge of the training/testing data:}
To consider a strong attack, we assume the attacker also has access to part of the training/testing data and the attacker has the capability to query sufficient data samples from the target model to perform MIAs including label-only attacks. The attack classifier is trained based on the output of these known data samples from the target model. This assumption allows more reasonable and realistic attacks and aligns with the state-of-the-art defense studies on MIAs~\cite{nasr2018machine,jia2019memguard,song2021systematic}.
In practice, if the attacker has less information, the attack performance would be weaker than what we have considered. 

\begin{figure}[!t]
\centering
\includegraphics[width=\linewidth]{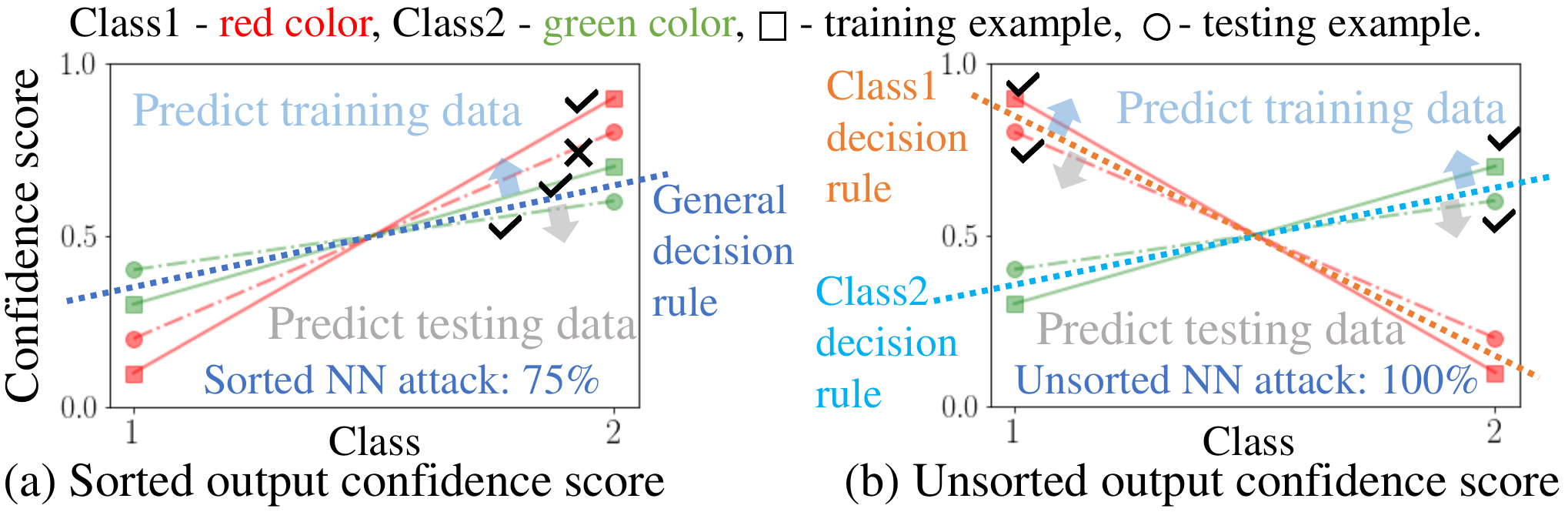}
\vspace{-15pt}
\caption{\small (a) Sorted, (b) Unsorted output confidence distribution for binary classification model. (Class1: $D_{train}$:[0.9, 0.8], $D_{test}$: [0.8, 0.2]; Class2: $D_{train}$:[0.3, 0.7], $D_{test}$: [0.4, 0.6]).}
\label{fig_nn_diff}
\vspace{-10pt}
\end{figure}

\vspace{-5pt}
\section{Design}
\label{design}

\subsection{Design overview}

To develop a defense that can fulfill all requirements in Sec.~\ref{mot}, it is important to understand the differences of MIAs and the impact of these differences on the defense.
In particular, we notice that the sorted and unsorted NN attacks learn different information from the different input formats.
The former learns \textbf{ONLY sorted} output confidence scores with label information eliminated \textit{(unified highest to lowest scores via sorting for any class)}, while the latter learns \textit{original unsorted scores together with label information}. The former relies on a general rule among all classes to infer members, thus attack accuracy is greatly impacted when the score difference between training and testing data \textit{at different classes} is different, as Fig.~\ref{fig_nn_diff}(a) shows. The latter learns a different output distribution \textbf{for each class} to attack like Fig.~\ref{fig_nn_diff}(b), however, incorrect prediction results introduce noises into attack training to reduce attack accuracy. 
Consequently, one attack is not uniformly stronger than the other, and it is necessary for a good defense to work against both.

\begin{figure*}[t]
\centering
\includegraphics[width=\textwidth]{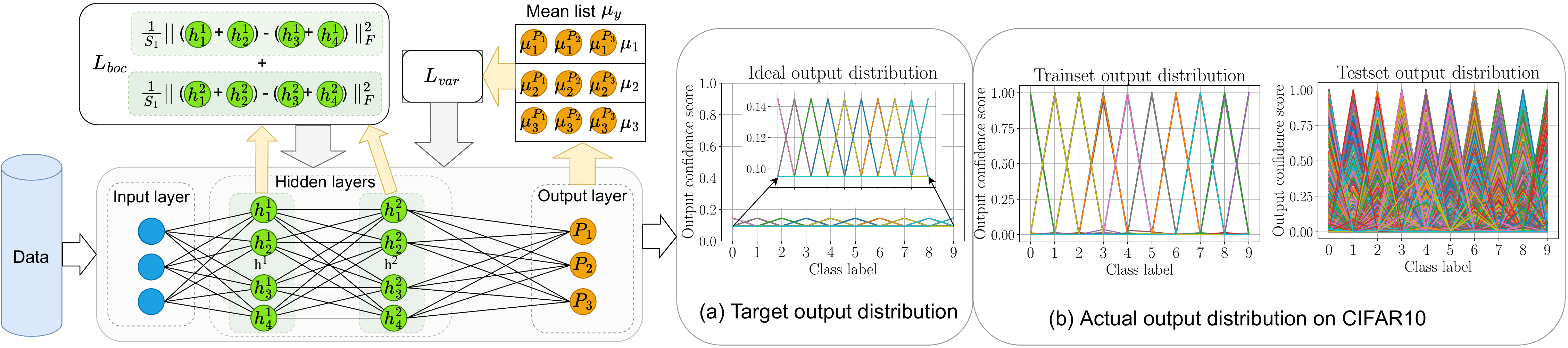}
\caption{\small An overview of our proposed \textbf{neuron guided} defense method with \textit{class-wise variance minimization} $L_{var}$ and \textit{layer-wise balanced output control} $L_{boc}$. 
(a) is the ideal output distribution that we aim to achieve by our proposed method (value space limited). (b) shows the actual output score distributions of training and testing set on the CIFAR10 by normal training.}
\label{fig_flow}
\vspace{-10pt}
\end{figure*}

With the understanding of the difference of MIAs, we now explore the ordinary case where the training accuracy and testing accuracy have a large gap, leading to a huge difference in {\it distributions} of confidence scores over the training and testing sets.
For example, Fig.~\ref{fig_flow}(b) shows the Alexnet model's output distribution on a real-world dataset CIFAR10 with 10 classes, where we can observe a substantial difference between the two output distributions of the training set and testing set.
It is indeed the difference between these distributions that is exploited by various attacks in different ways to infer membership in the training set.
Existing defenses either propose to add an adversarial regularization term during the training or to manipulate the model's outputs during the inference stage to reduce the distribution gap. 
However, they either cannot provide enough distortions in training to suppress the difference (see Fig.~\ref{fig_output_dis}(b)) or can only defend against one type of attacks (see Fig.~\ref{fig_output_dis}(a), where the perturbed output cannot defend against unsorted NN attack, because the false predicted testing set can be easily distinguished by the attack classifier). 

In summary, we can learn that \textbf{for models with overfitting and a large accuracy gap between training and testing data, only a uniform distribution with marginal difference between classes can deceive both sorted and unsorted NN attacks and serve as an effective distribution against MIAs.}
One possible path to this end is to restrict all confidence scores to be evenly distributed and thus reduce the output distribution gap between the training set and testing set. As the example in Fig.~\ref{fig_flow}(a) shows, in a 10 class output, when all of the confidence scores of each class are close, the value of each confidence score should be close to the mean value (e.g. 0.1), where the score of correct label is slightly higher than the mean value and that of others are lower than it. 
Note that while it is not possible to ``know" the distribution of scores for the testing set, limiting the distribution space of the model output is an indirect way to ensure closeness of the distributions arising from training and testing sets.

To achieve this goal,
we propose a novel framework to control the output confidence score through dedicated output and inner (hidden) neuron control.
As Fig.~\ref{fig_flow} shows, we seamlessly integrate our class-wise variance minimization (for output neurons) into the customized layer-wise balanced output control (for hidden neurons) intermediate results to construct the privacy preserving neural network model. 
In the following,
we will first 
detail the 
class-wise variance minimization regularization, then
the layer-wise balanced output control regularization, and finally present the combined training process to generate defense-efficient outputs.

\subsection{Class-wise variance minimization}
We propose to enforce the output confidence scores of all classes close to their mean distribution during the training.
In this way, we expect that the output distribution can be in a similar range 
in both the training and testing set.
In particular, we choose to add the class-wise variance as a regularization term $L_{var}$ into the loss function.
Since variance directly  measures the spread around the mean value, minimizing the variance allows us to control
the space of output distributions such that the training and test output distributions are close to an “ideal”
distribution.
\begin{equation}\label{eqn_1}
\small
L_{var} =  \frac{1}{N} \textstyle\sum_{i=0}^{N} (F(x)-\mu_y )^{2} 
\end{equation}
where $F(x)$ is the target model output confidence score for the training data $(x,y)$, and $N$ is the batch size. 
$ \mu_y $ represents the mean list of corresponding class $y$ that is used to calculate the expectation of the squared deviation of the output. For example, as Fig.~\ref{fig_flow} shows, the mean list $\mu_1$ is a three dimensional vector--($\mu_1^{P_1},\mu_1^{P_2}, \mu_1^{P_3}$) corresponding to three output neurons, which is an average output confidence score vector for data samples with the class label 1.

There are different ways to calculate the mean vector $\mu$ for the  variance minimization to train neural networks. 
In the following we
discuss three of them and provide a reasoning for choosing the class-wise approach.
We briefly show the model accuracy and the neural network based MIA of CIFAR10 classification task with three different variance calculation methods on Table~\ref{var_diff}.

\noindent \textbf{Batch-wise:} 
The simplest way is to perform the batch-wise variance calculation that minimizes the variance of each batch's output.
The mean list calculated in each batch can have different results
and the variance minimization may not have a consistent goal.
Thus, this method can maintain the model accuracy at the standard level, but it does not provide the necessary defense.

\noindent \textbf{Single-sort:} 
Another way is to calculate variance with a single sorted mean list, which means to sort all outputs and maintain a sorted mean list for all data to calculate the variance of each output. 
This approach provides a uniform goal unlike the batch method, however the distribution and accuracy for each class is highly variable.
As a result, training with this regularization term cannot converge and leads to a huge model accuracy loss. 

\noindent \textbf{Class-wise:} In order to set a consistent goal for the variance minimization as well as consider the effects of each class, we proposed a fine-tuned approach to calculate the output variance across multiple predictions from the same class via updating the class-wise mean list.
Since we calculate the empirical mean of the output confidence scores for each class, we do not need to sort the output confidence scores and can learn more information from the relationship of each label in the output distribution.
Table~\ref{var_diff} shows that the class-wise variance minimization can provide best utility-defense trade-off.

While class-wise variance addresses the output distribution difference between training and testing set to an extent, weakness of existing defenses against a sorted NN attack suggests that a finer control of model training is required for which we propose the layer-wise output control, described in the subsequent section.

\begin{table}[]
\caption{\small Testing/MI accuracy on unsorted NSH and sorted NN model with variance minimization training on CIFAR10.}
\small
\vspace{-5pt}
\centering
\begin{tabular}{|c|c|c|c|c|}
\hline
\multicolumn{2}{|c|}{$\beta=300$ }   & Batch-wise & Single-sort & Class-wise \\ \hline \hline
\multicolumn{2}{|c|}{{Training set accuracy}}    & 74.09     & 51.77       & 71.29      \\ \hline
\multirow{2}{*}{{\begin{tabular}[c]{@{}c@{}}MI \\ accuracy\end{tabular}}} & {Sorted NN}   & 71.41     & 60.18       & 50.4       \\ \cline{2-5} 
& {Unsorted NSH}          & 62.39     & 55.74       & 57.41      \\ \hline
\end{tabular}
\label{var_diff}
\vspace{-10pt}
\end{table}

\subsection{NN layer-wise balanced output control}

In this section, we propose 
to perform 
a neuron regularization through layer-wise balanced output control in the training process to further constrain the output distribution.
The neuron regularization is first introduced to determine the weight values in favor of the pruning goals during the training process~\cite{zhuang2020neuron, wang2020neural}. 
Unlike their design to manipulate the parameters to 0 or some specific values, our design focuses on balancing the parameters ~\cite{ozen2020concurrent} in each layer so that there will be no particular parameters to dominate the prediction results.
To perform the layer-wise balanced output control, we separate the output of each layer into two groups, calculate the summation within each group and finally add the mean square error for the group output difference at each layer. The final summation $L_{boc}$ is as a regularization term and added to the loss function for  training.
\begin{equation}
\label{eqn_2}
L_{boc} = \textstyle \sum_{l=1}^{M} \frac{1}{S_{l}}\left\|\textstyle\sum_{i=1}^{\left\lfloor S_{l} / 2\right\rfloor} h_{i}^{l}-\textstyle\sum_{i=\left\lfloor S_{l} / 2\right\rfloor+1}^{S_{l}} h_{ i}^{l}\right\|_{F}^{2}
\end{equation}

In the equation, $M$ is the number of layers, and $S_l$ is the number of layer $l$'s outputs. $h_{i}^{l}$ denotes the $i$'s output on layer $l$. The balanced output control on each layer can help to regularize the effects of individual outputs in a general sense as we try to minimize the difference of the output groups.
Thus there will be no extremely large values and the intermediate results will be more balanced.
Combining this layer-wise balanced output control with variance minimization method for the model training,
we can train the target model with high utility and defense effects to the best trade-off.

\subsection{Neuron regularization-based training flow }
\label{reg-method}
With two regularization terms used for the training, our overall loss function is as follows:
\begin{equation}\label{reg}
Loss = L(F) + \alpha \cdot L_{boc} + \beta \cdot L_{var},  
\end{equation}
where we use hyperparameter $\alpha$ and $\beta$ to control the balance between the optimizing classification task and the effects to constrain the output distribution to perform the defense.
Here we present the loss calculation algorithm that contains both 
class-wise variance minimization and layer-wise balanced output control.
During each batch of training, the training algorithm will first calculate the difference of two evenly split groups in each layer $l$ using the intermediate results $h^{l} $ to compute $L_{boc}$. 
Then it will update the class-wise mean list $\mu_y$ with all data records. With the updated $\mu_y$, it can calculate the variance $L_{var}$ for the data.
Finally, the total loss can be added and used for the weight update.
The detailed pseudo code is shown in Algorithm \ref{alg_1} in Appendix.

Using our proposed regularization method, we can further improve the performance by amplifying the layer-wise intermediate features. In some cases such as convolutional layers, given the complex feature extraction from coarse to fine through a layer-wise manner,  
the original feature maps may not be salient enough for our proposed method to learn during the training. Thus in practice, we can perform a layer-wise amplification to intentionally enlarge the top $\eta \%$ of the feature maps' values 
in the training process, so that the proposed solution in Eqn.~\ref{reg} can better learn and regularize the most important features, making the model output scores converge to desired distributions easily.
Furthermore, we could also leverage 
such layer-wise amplification during the inference to alleviate the effect of amplified features by training for better maintaining the model utility. We demonstrate its effectiveness in Appendix~\ref{eff_layer_amp}.

\vspace{-3pt}
\section{Experimental setup}
\noindent {\bf Datasets and parameter setting.}
We use three benchmark datasets Texas100 ~\cite{shokri2017membership}, CIFAR10 and CIFAR100 ~\cite{krizhevsky2009learning} to demonstrate the effectiveness of our \textit{NeuGuard} defense against membership inference attack for different application scenarios. We follow the data splitting strategy in ~\cite{nasr2018machine,jia2019memguard} (See Table~\ref{data_size}) and have the detailed description in the Appendix \ref{datasets}.

We compared our method  with three state-of-the-art 
MI defense methods, i.e., training time based adversarial regularization (AdvReg)~\cite{nasr2018machine}, Early stopping~\cite{song2021systematic}, and inference time based MemGuard~\cite{jia2019memguard}. We also select the regular
model without defense for comparison. 
We use the publicly available source code
of the three methods 
for evaluation.  
All methods are implemented in Pytorch~\cite{paszke2019pytorch}. 
All experiments are run on a linux PC with AMD Ryzen Threadripper 2990WX 32-Core Processor, 128 GB memory and NVIDIA GeForce RTX 2080 Ti GPU with 11 GB graphic memory.

We follow the training method and hyperparameter setup on each defense methods proposed by the authors and use the published code to evaluate the corresponding defense mechanism. We choose the best results to compare in the evaluation. The detailed setting can be found in Appendix~\ref{params}.

As for our proposed method, we apply the \textit{NeuGuard} methods with different settings.
For the fully connected model on the Texas100 dataset, we apply the variance minimization method with $\beta=3000$ and layer-wise balanced output control with $\alpha=100$. 
For CIFAR100 task, we set the $\beta=1000$ and $\alpha=5$. For the proposed methods to better learn and control the output, we amplify the top 10\% of the convolution layer's feature map 2 times during the training, and amplify 2 times of the top 25\% feature map in the inference stage. 
For CIFAR10 task, we set the $\beta=200$, $\alpha=30$, and amplify 1.5 times the top 10\% feature map in training. For the inference stage, we amplify 1.5 times of the top 35\% feature map to 
obtain the best utility-privacy trade-off. 

\noindent {\bf Attack setup.}
We consider the existing two types of attack models: neural network (NN) based attack and metric based attack.
We follow the model structure and standard setup in ~\cite{nasr2018machine,jia2019memguard} that use different fully connected neural network as attack classifier. The sorted NN attack classifier contains three hidden layers with 512,256 and 128 neurons, respectively. The unsorted NSH attack classifier consist of three neural networks as introduced in Section \ref{intro_nn_MIA}.
Please also see the detailed description in Appendix~\ref{nn_attack_setup}.

The metric-based MIAs use the output confidence score vector to calculate metrics and compare the results with preset thresholds to determine the prediction.
We evaluate our work with four state-of-the-art 
metrics for MIAs following the approach proposed in ~\cite{song2021systematic}, which includes
prediction correctness, prediction confidence, prediction entropy and modified prediction entropy (see Section \ref{metricbased_attack}). 
Moreover, we follow the strongest label attack from ~\cite{choquette2021label}
that use C$\&$W attack to generate adversarial examples to further demonstrate the effectiveness of our defense method.

In our evaluation, we follow 
~\cite{nasr2018machine,nasr2019comprehensive}, which implies a strong adversary that knows a substantial part of the training set and will use it to train the inference attack models. 
In particular, we sample the input data from training set and testing set with an equal 0.5 probability following prior works
~\cite{shokri2017membership,song2019privacy,yeom2018privacy} to maximize the uncertainty of membership inference attacks. In this way, we can make the attack accuracy unbiased and easy to analyze.

\noindent {\bf Evaluation metrics.}
We use three metrics to evaluate defenses.
\begin{itemize}[leftmargin=*]
\item \textbf{Membership inference (MI) accuracy:} 
It is the accuracy of an attack to infer the membership of data in the training set.
A good defense should 
lead the membership inference attack accuracy to a random guess (e.g. $\sim50\%$).

\item \textbf{Testing accuracy:} 
It is the accuracy of a defense method on the testing set.
A good defense should obtain the testing accuracy close to the trained model without defense.

\item \textbf{Running time:}
We measure the model training time and inference time of the defense methods and set the time used for a regular model as the baseline. A good defense should not cause large overhead compare to the baseline.

\end{itemize}


\begin{table}[!t]
\caption{\small Results of compared defenses against NN based MI attacks. AdvReg and MemGuard are originally designed to defend against sorted NN and unsorted NSH attacks, respectively. Baseline is the normal training without defense. }
\label{tbl:nn_attack}
\center
\resizebox{\linewidth}{!}{%
\begin{tabular}{|c|c|c|c|c|c|c|}
\hline
\multicolumn{2}{|c|}{\textbf{Texas100}} &
  Baseline &
  Early stopping &
  AdvReg &
  MemGuard &
  \textbf{\textit{NeuGuard}} \\ \hline \hline
\multicolumn{2}{|c|}{Testing   accuracy}    & 58.5  & 50.9  & 51.2   & 58.5  & 55.8  \\ \hline
\multirow{2}{*}{\begin{tabular}[c]{@{}c@{}}MI \\ accuracy\end{tabular}} & Unsorted NSH         & 65.75 & 57.42 & 64.18  & 50.83 & 50.58 \\ \cline{2-7} 
& Sorted NN            & 60.98 & 53.32 & 53.48  & 60.52 & 54.54 \\ \hline\hline
\multicolumn{2}{|c|}{Training time(s)}   & 0.006 & 0.006 & 0.328 & 0.006 & 0.045 \\ \hline
\multicolumn{2}{|c|}{Training overhead}  & 1$\times$    & 1$\times$     & 54.7$\times$   & 1$\times$     & 7.5$\times$  \\ \hline
\multicolumn{2}{|c|}{Inference time(s)}  & 0.002 & 0.002 & 0.002  & 1.8   & 0.002 \\ \hline
\multicolumn{2}{|c|}{Inference overhead} & 1$\times$    & 1$\times$    & 1$\times$     & 900$\times$  & 1$\times$    \\ \hline
\end{tabular}
}
\center
\resizebox{\linewidth}{!}{%
\begin{tabular}{|c|c|c|c|c|c|c|}
\hline
\multicolumn{2}{|c|}{\textbf{CIFAR100}} &
  Baseline &
  Early stopping &
  \begin{tabular}[c]{@{}c@{}}AdvReg\end{tabular} &
  \begin{tabular}[c]{@{}c@{}}MemGuard \end{tabular} &
  \textbf{\textit{NeuGuard}} \\ \hline\hline
\multicolumn{2}{|c|}{Testing accuracy}      & 43.8  & 41.0    & 39.6    & 42.9  & 43.0     \\ \hline
\multirow{2}{*}{\begin{tabular}[c]{@{}c@{}}MI \\ accuracy\end{tabular}} & Unsorted NSH         & 80.95 & 60.70  & 62.67   & 50.41 & 51.42    \\ \cline{2-7}
& Sorted NN            & 81.42 & 59.62 & 58.64   & 59.63 & 57.82  \\ \hline\hline
\multicolumn{2}{|c|}{Training time(s)}   & 0.017 & 0.017 & 0.050 & 0.017 & 0.045 \\ \hline
\multicolumn{2}{|c|}{Training overhead}  & 1$\times$    & 1$\times$    & 2.96$\times$   & 1$\times$     & 2.62$\times$  \\ \hline
\multicolumn{2}{|c|}{Inference time(s)}  & 0.017 & 0.017 & 0.017   & 1.7   & 0.025  \\ \hline
\multicolumn{2}{|c|}{Inference overhead} & 1$\times$    & 1$\times$    & 1$\times$      & 100$\times$  & 1.47$\times$  \\ \hline
\end{tabular}
}

\center
\resizebox{\linewidth}{!}{%
\begin{tabular}{|c|c|c|c|c|c|c|}
\hline
\multicolumn{2}{|c|}{\textbf{CIFAR10}} &
  Baseline &
  Early stopping &
  \begin{tabular}[c]{@{}c@{}}AdvReg\end{tabular} &
  \begin{tabular}[c]{@{}c@{}}MemGuard \end{tabular} &
  \textbf{\textit{NeuGuard}} \\ \hline\hline
\multicolumn{2}{|c|}{Testing   accuracy}      & 76.6  & 71.1  & 71.1  & 76.6  & 74.6  \\ \hline
\multirow{2}{*}{\begin{tabular}[c]{@{}c@{}}MI \\ accuracy\end{tabular}} & Unsorted NSH         & 71.70  & 60.07 & 61.20  & 51.43 & 51.57 \\ \cline{2-7} 
& Sorted NN            & 70.59 & 57.47 & 56.18 & 62.73 & 55.60  \\ \hline \hline
\multicolumn{2}{|c|}{Training time(s)}   & 0.017 & 0.017 & 0.050 & 0.017 & 0.046 \\ \hline
\multicolumn{2}{|c|}{Training overhead}  & 1$\times$    & 1$\times$    & 2.94$\times$ & 1$\times$    & 2.71$\times$ \\ \hline
\multicolumn{2}{|c|}{Inference time(s)}  & 0.017 & 0.017 & 0.017 & 1.7   & 0.027 \\ \hline
\multicolumn{2}{|c|}{Inference overhead} & 1$\times$    & 1$\times$    & 1$\times$    & 100$\times$  & 1.59$\times$ \\ \hline
\end{tabular}
}
\vspace{-10pt}
\end{table}

\begin{figure}[!t]
\centering
\includegraphics[width=\linewidth]{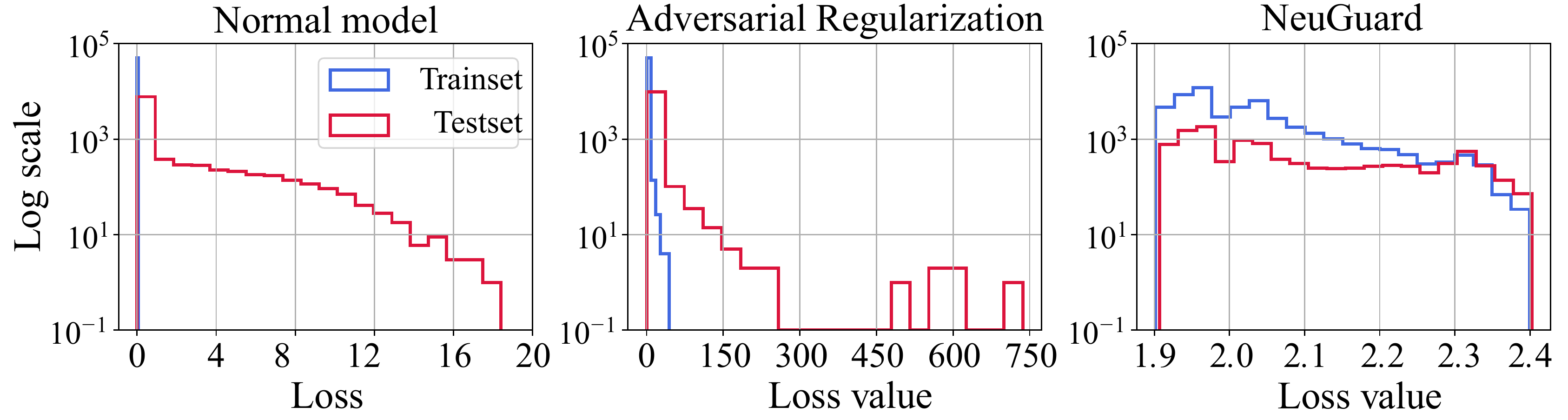}
\caption{\small Loss distribution on CIFAR10 with regular training, adversarial regularization training, and our \textit{NeuGuard}.} 
\label{fig_loss_dis}
\vspace{-10pt}
\end{figure}

\section{evaluation on nn based attacks}
\label{nn_based_attack_evaluation}
In this section, we evaluate our proposed defense against  neural network based MIAs (see Section~\ref{intro_nn_MIA}). For comprehensive evaluations, we consider the existing two well-known attack models: \textit{sorted attack} and \textit{unsorted attack}.

\subsection{Experimental results comparison}
\label{comb_res}
\begin{figure*}[!t]
\centering
\subfigure[\small MemGuard]{%
\begin{minipage}[t]{0.25\linewidth}
\centering
\includegraphics[width=\textwidth]{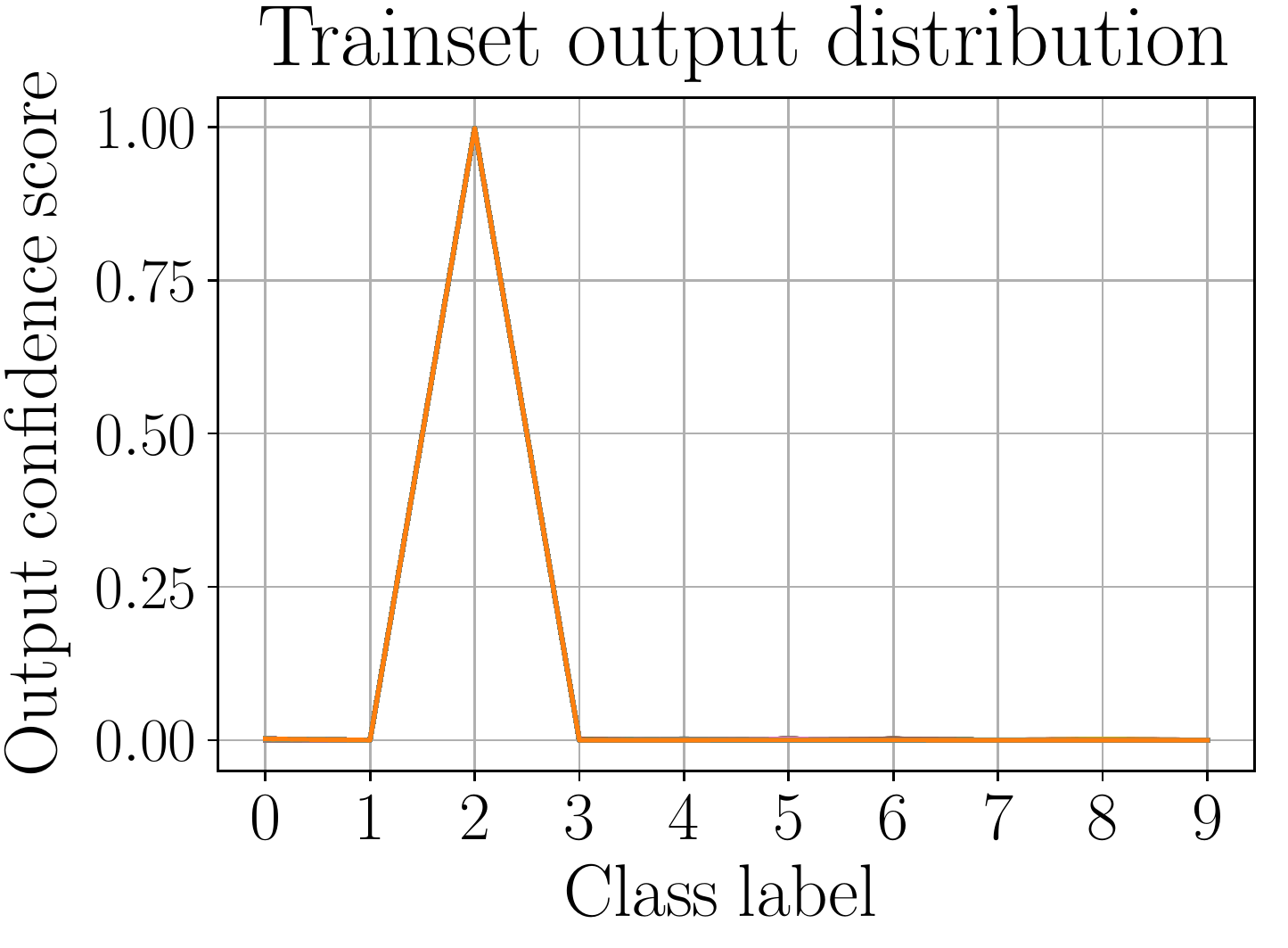}
\includegraphics[width=\textwidth]{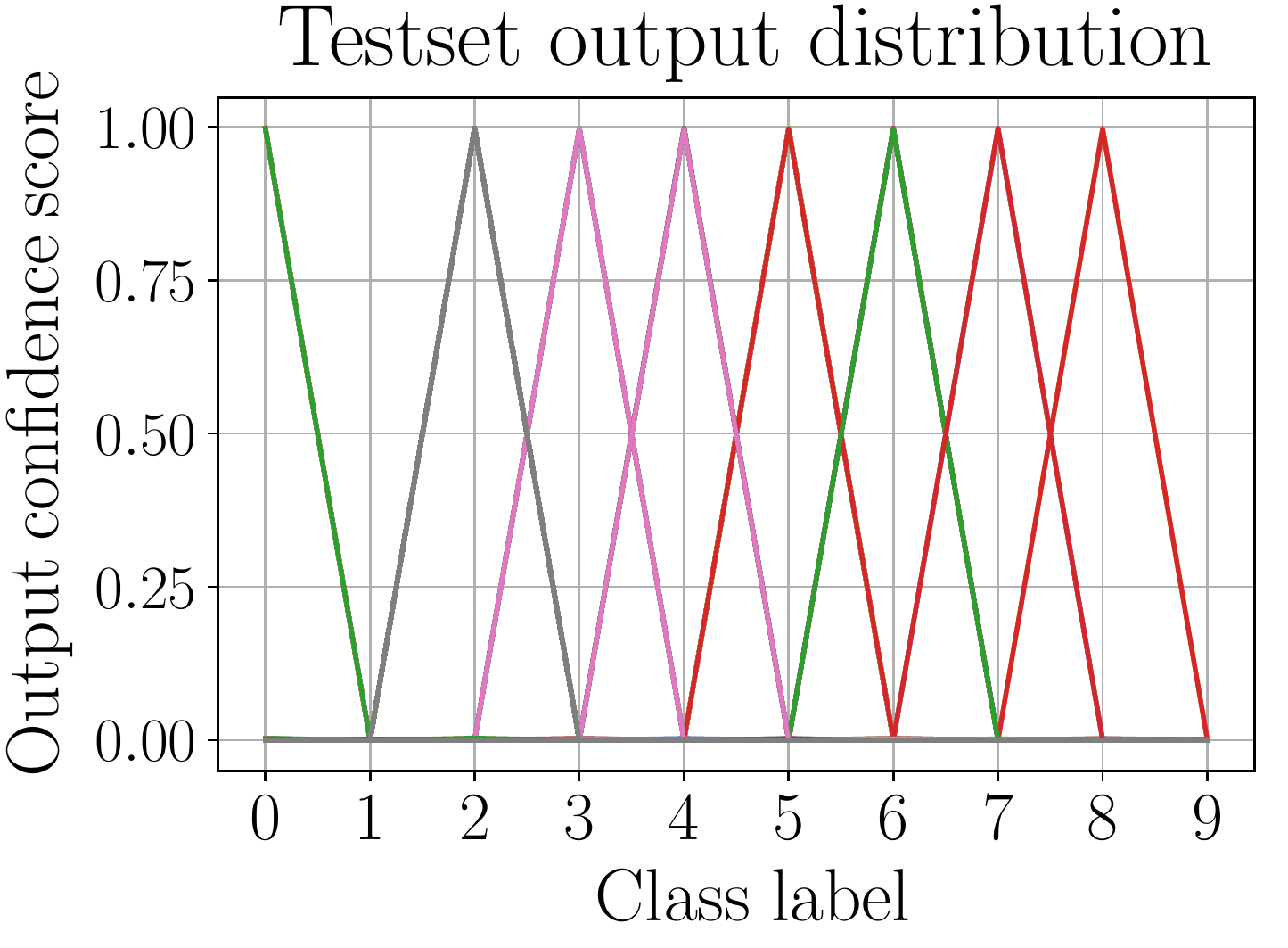}
\end{minipage}
}
\quad
\subfigure[\small AdvReg]{%
\begin{minipage}[t]{0.25\linewidth}
\centering
\includegraphics[width=\textwidth]{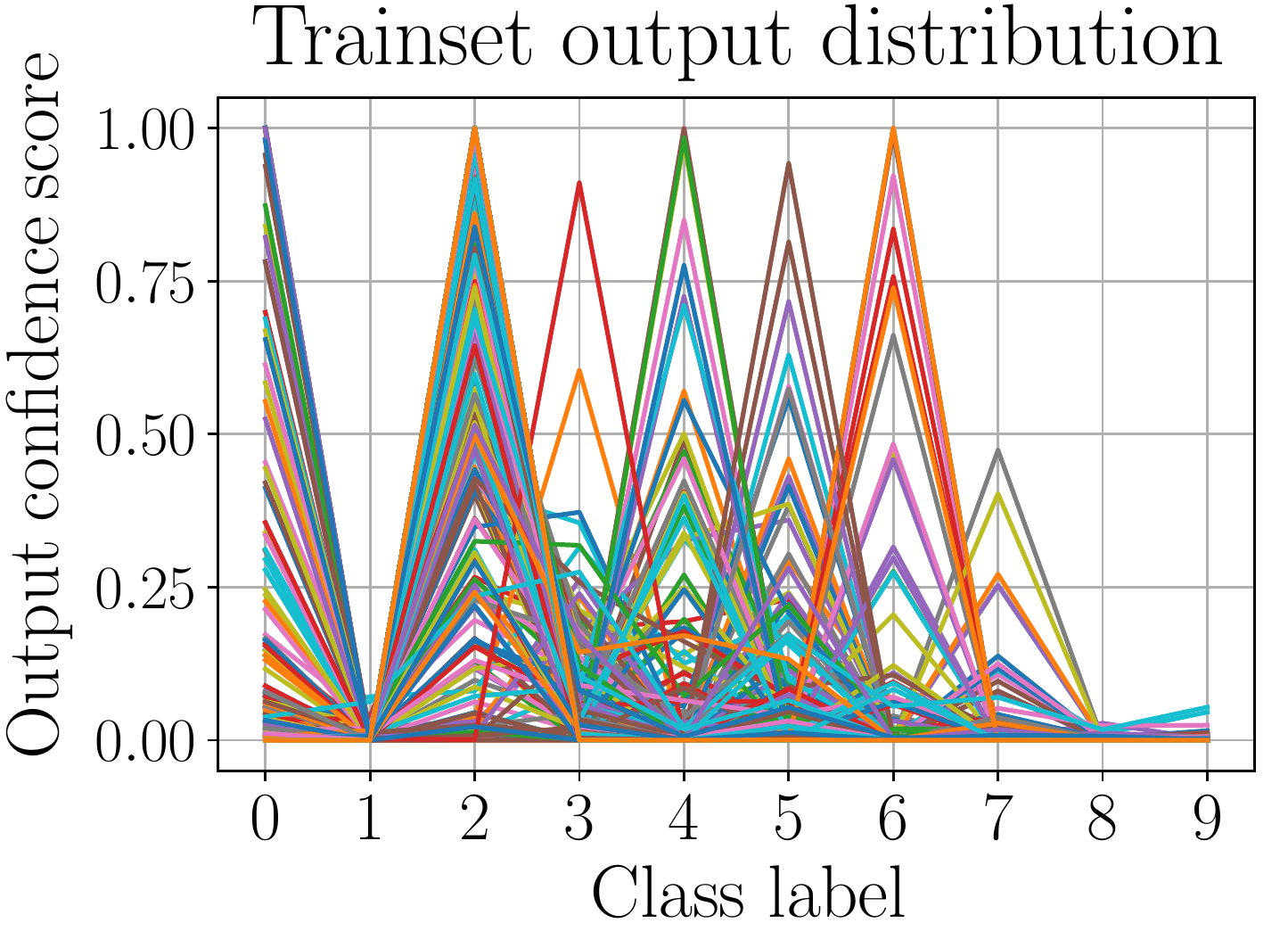}
\includegraphics[width=\textwidth]{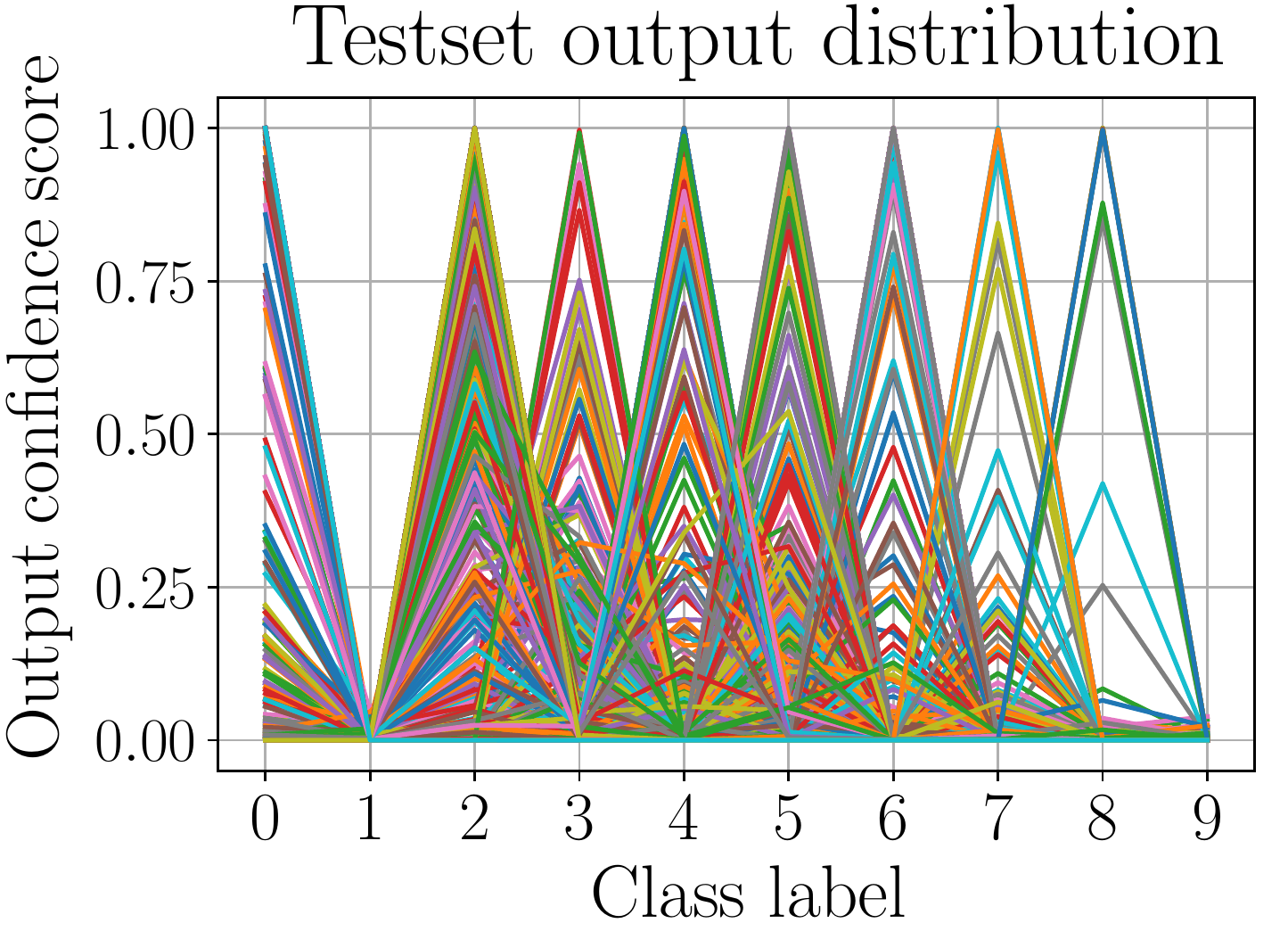}
\end{minipage}
}
\quad
\subfigure[\small \textit{NeuGuard}]{%
\begin{minipage}[t]{0.25\linewidth}
\centering
\includegraphics[width=\textwidth]{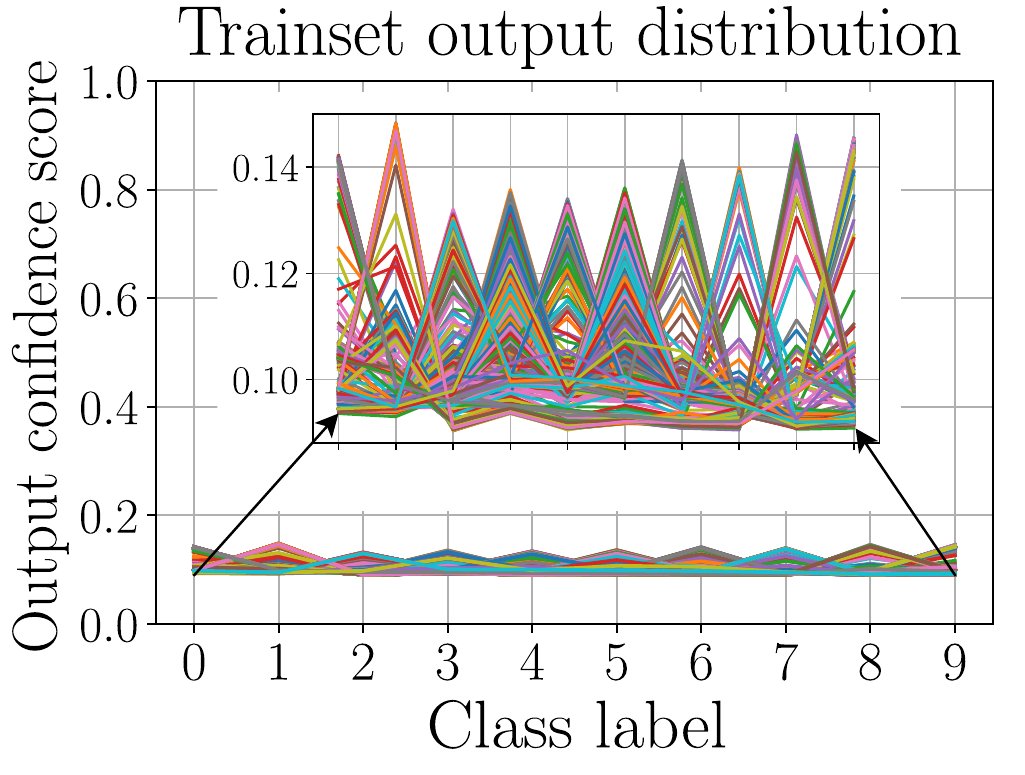}
\includegraphics[width=\textwidth]{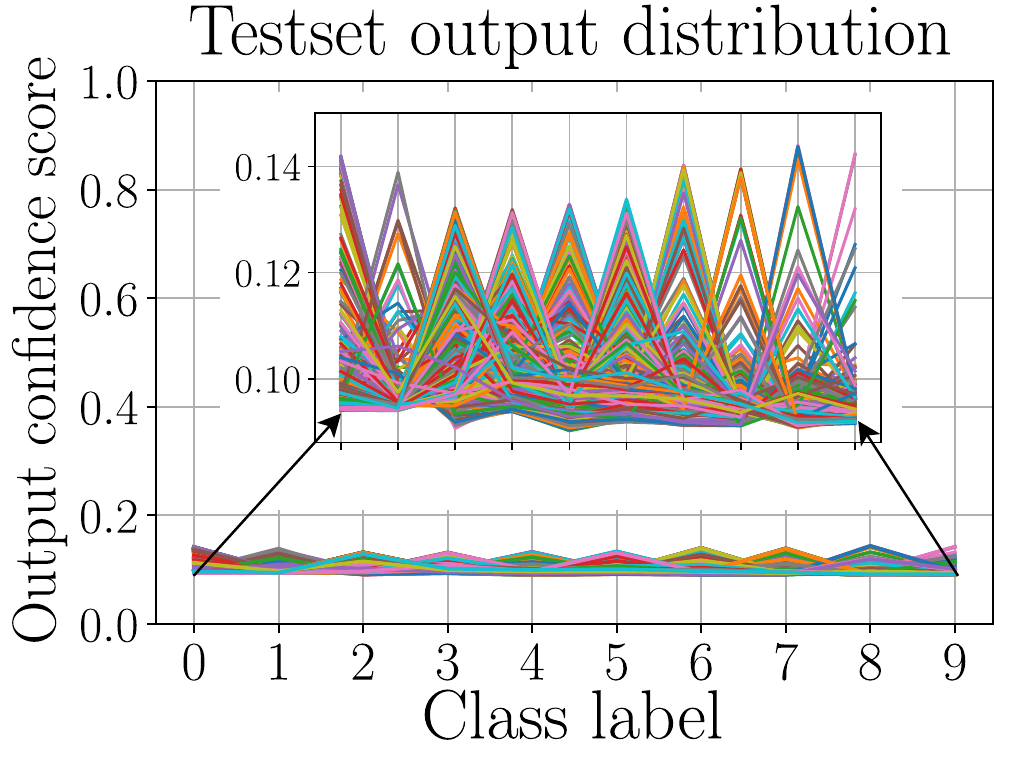}
\end{minipage}
}
\vspace{-10pt}
\caption{ Output distribution of training and testing samples with class ``2" in CIFAR10 for the compared defenses. Each color line represents a sample's output confidence scores.}
\label{fig_output_dis}
\vspace{-10pt}
\end{figure*}

Table~\ref{tbl:nn_attack} shows the testing accuracy, MI accuracy, and running time of the compared defenses on the three benchmark datasets.
We have the following observations: 

\noindent \textbf{Our \textit{NeuGuard} achieves the best utility-privacy trade-off against both the sorted and unsorted attacks among evaluated solutions, indicating by far the best defense effectiveness and generality while maximizing the achievable utility.} 
As Table~\ref{tbl:nn_attack} shows, \textit{NeuGuard} effectively mitigates both the unsorted NSH attack and sorted NN attack.
Specifically, with \textit{NeuGuard}, the unsorted attack only achieves an MI accuracy close to random guessing on all the three datasets. For the sorted NN attack, \textit{NeuGuard} can significantly decrease the MI accuracy from 60.98\%, 81.42\%, and 70.59\% to 54.54\%, 57.82\% and 55.6\% on the three datasets, respectively. Moreover, the testing accuracy obtained by \textit{NeuGuard} is close to those obtained by the Baseline method without defense. 
We also observe that while \textbf{MemGuard} can maintain the same testing accuracy, it cannot defend against both kinds of NN based attacks simultaneously. To comprehensively compare MemGuard with our defense approach, \textbf{1)} we obtain its defense results \ul{using an unsorted NSH attack classifier as the target defense model} to generate the noised output confidence scores. As expected, MemGuard achieves comparable results as our \textit{NeuGuard} in defending against the unsorted NSH attack.
However, it is not effective enough against the sorted NN attack, e.g., $60\%$ MIA accuracy on all the three datasets. \textbf{2)} For the original design \ul{using a sorted NN attack classifier as the defense model}, MemGuard exhibits a similar behavior: defending against sorted NN attack to $\sim50\%$ but $>60\%$ for unsorted NSH attack. We further show the output confidence scores distribution in this case in Fig.~\ref{fig_output_dis}(a) and analyze the reasons why this is not a general defense in the following section.
\textbf{AdvReg and Early stopping} show similar defense effects on both attacks in most cases, which is also verified in ~\cite{song2021systematic}. 
Comparing with \textit{NeuGuard}, they achieve the similar level of defense effectiveness against the sorted NN attack, but perform much worse against the unsorted NSH attack on the three datasets. 
In addition, they also incur a non-negligible (more prominent) utility loss, i.e., large testing accuracy gaps between them and the Baseline method. 

\noindent \textbf{Our \textit{NeuGuard} has smaller overhead and better performance.}
For the training and inference overhead, we compare our method with AdvReg and MemGuard separately. In Table~\ref{tbl:nn_attack} we calibrate the training and inference time for processing a single batch of data and use the regular training and inference time as the baselines for overhead comparisons.
As the data loader used to load each batch of images dominates the processing time~\cite{yang2019accelerating}, the training and inference time on CIFAR100 and CIFAR10 are the same in the table.
Specifically, 
since we apply layer-wise amplification in both training and inference stages, this incurs some overheads.

In the fully connected neural network case on the Texas100 dataset,
our \textit{NeuGuard} training method has $\sim7.5\times$ overhead while AdvReg training method causes $\sim54.7\times$ overhead compared to the regular training. 
As for the inference stage, the MemGuard method causes $\sim900\times$ overhead as it might need to run several hundred times of inference to search the noisy output confidence vector satisfying their requirements. Our methods, however, leads to no overhead since we have no additional operation in the inference. 

For convolutional neural network models, the training overhead incurred by our defense approach and the other approaches are all fairly close to each other.
The overhead caused by our method in the inference stage is much less than the MemGuard method (1.47$\times$ and 1.59$\times$ vs. 100$\times$ and 100$\times$ in CIFAR100 and CIFAR10 cases). This is mainly because our layer-wise amplification is a simple one-time process added in the inference stage.
In conclusion, our defense can achieve the best defense effectiveness with a marginal model accuracy drop. The overheads caused in training and inference are relatively low compared to other defense mechanisms.

\vspace{-10pt}
\subsection{Why does \textit{NeuGuard} perform better?}
One key reason that our defense obtains the best utility-privacy trade-off is because it generates the output confidence scores of the training set and testing set with the smallest variance. 
Table~\ref{mean_var_cifar100} shows the variance of the output confidence scores calculated on the training set and testing set for CIFAR100 with different methods. We observe the similar results for Texas100 and CIFAR10.

\noindent \textbf{\textit{NeuGuard} obtains the smallest variance of the output confidence scores.} 
Our defense reduces the variance of the output confidence score by three orders of magnitude compared to the baseline model, while all other defense methods keep it at the same level. Our method also decreases the variance gap between the training set and testing set
(see Table~\ref{mean_var_cifar100} in Appendix). 
The results show \textit{NeuGuard} can effectively suppress final outputs' variance. 

\noindent \textbf{\textit{NeuGuard} delivers the most consistent loss distribution between the training set and testing set.}  
Fig.~\ref{fig_loss_dis} illustrates the loss distributions of the training set and testing set for CIFAR10 under different training methods. Here we use CIFAR10 as an example, since we observe the similar trend on Texas100 and CIFAR100. 
With the normal training in the baseline method, the training set accuracy will reach $\sim100\%$ when model is overfitted and the loss distribution ranges from 0 to 0.07 while that of the testing set ranges from 0 to 18.  AdvReg attempts to reduce the gap between training set and testing set by reducing the training set accuracy and leading the loss distribution range to [0, 50] for the training set. However, the noise introduced by  AdvReg also causes a test accuracy drop and significantly increases the range of the testing set loss to [0, 750]. 
In contrast, our \textit{NeuGuard} constructs two similar loss distributions for both the training set and the testing set, and restricts them to a very small range--[1.9, 2.5].

\begin{figure}[t]
\centering
\includegraphics[width=\linewidth]{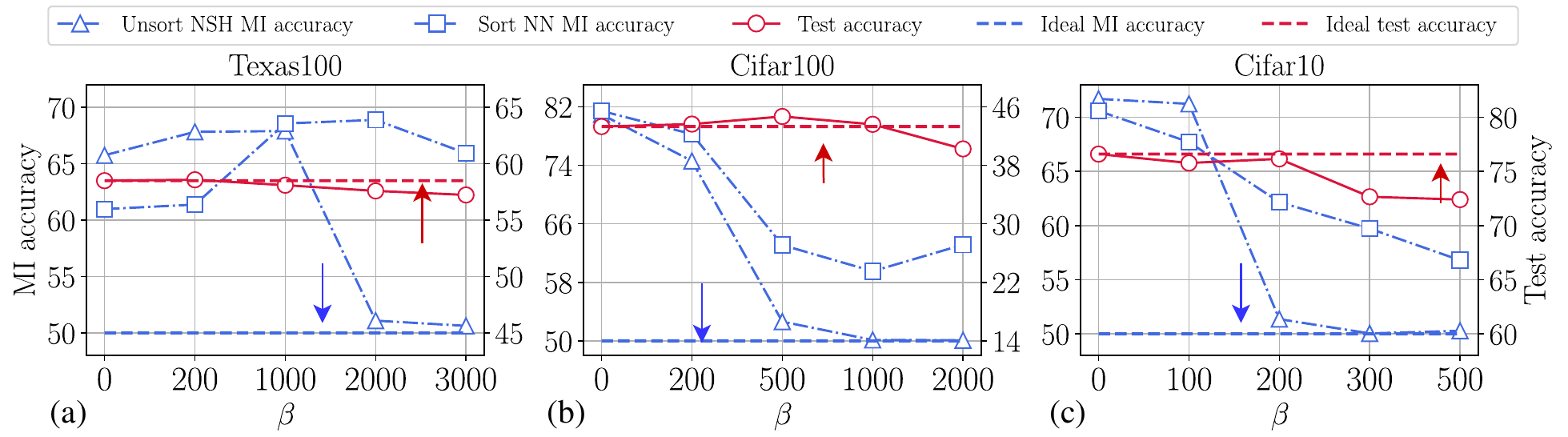}
\caption{\small 
Test accuracy and NN based MI accuracy with different $\beta$ for class-wise variance minimization control term. (a) Texas100, (b) CIFAR10, and (c) CIFAR100.
}
\label{fig_min_var_beta}
\vspace{-10pt}
\end{figure}

\noindent \textbf{Visualizing data samples' output confidence scores.} To better understand the different defense results among the compared defenses, we further visualize the output confidence score vectors of the training and testing samples generated by these defenses. 
Specifically, we randomly choose a class (class ``2" in our experiment) from CIFAR10 and show the output confidence score vectors of 1000 training and testing samples each from this class in Fig.~\ref{fig_output_dis}. Note that each color line corresponds to one sample's output confidence score vector (each has ten scores for ten classes). \ul{We have the following observations:}

\ul{First}, MemGuard \ul{with the sorted NN attack classifier as defense model} can construct two similar output distributions for the training set and testing set, i.e., the output confidence score vector has one extremely large score for one class and almost equally small scores for all the other classes and most of the outputs overlap as Fig.~\ref{fig_output_dis} (a) shows. In this case, by sorting the confidence score vectors for training 
and testing samples, the output distributions of training set and testing set are almost the same. 
This observation explains why MemGuard in this case can defend against the sorted attacks. 
However, without sorting, the two output distributions are unlike, i.e, the classes associated with the largest output confidence scores are different (see Fig.~\ref{fig_output_dis} (a)) and the unsorted attacks can capture this information to perform the attack. Thus, MemGuard cannot effectively defend against unsorted attacks.

\ul{Second}, AdvReg perturbs the training samples during training and attempts to make the training set output distribution similar to the testing set output distribution. However, as we observe from Fig.~\ref{fig_output_dis} (b), 
the two distributions are not close enough. Therefore, besides suffering from testing accuracy loss, AdvReg neither exhibits strong defense effectiveness.

\ul{Third}, unlike MemGuard and AdvReg, with the class-wise variance minimization and layer-wise balanced output control, \textit{NeuGuard} controls and orchestrates the final output neurons' results and intermediate neurons' results for a destined output (confidence score) distribution. 
As a result, \textit{NeuGuard} can generate the targeted output confidence score vectors for all training and testing samples, where all values in the score vector are close to the mean value, i.e., one over the number of total classes, as shown in Fig.~\ref{fig_output_dis}(c).
Such results have two important implications:
i) The output confidence scores of member and non-member training samples are similar, making \textit{NeuGuard} effective to defend against (both sorted and unsorted) membership inference attacks; 
ii) The output distributions of training samples and testing samples are very close, meaning that \textit{NeuGuard} has good generalization ability. 
Therefore, \textit{NeuGuard} obtains the best utility-privacy trade-off. 

\begin{figure}[!t]
\centering
\includegraphics[width=\linewidth]{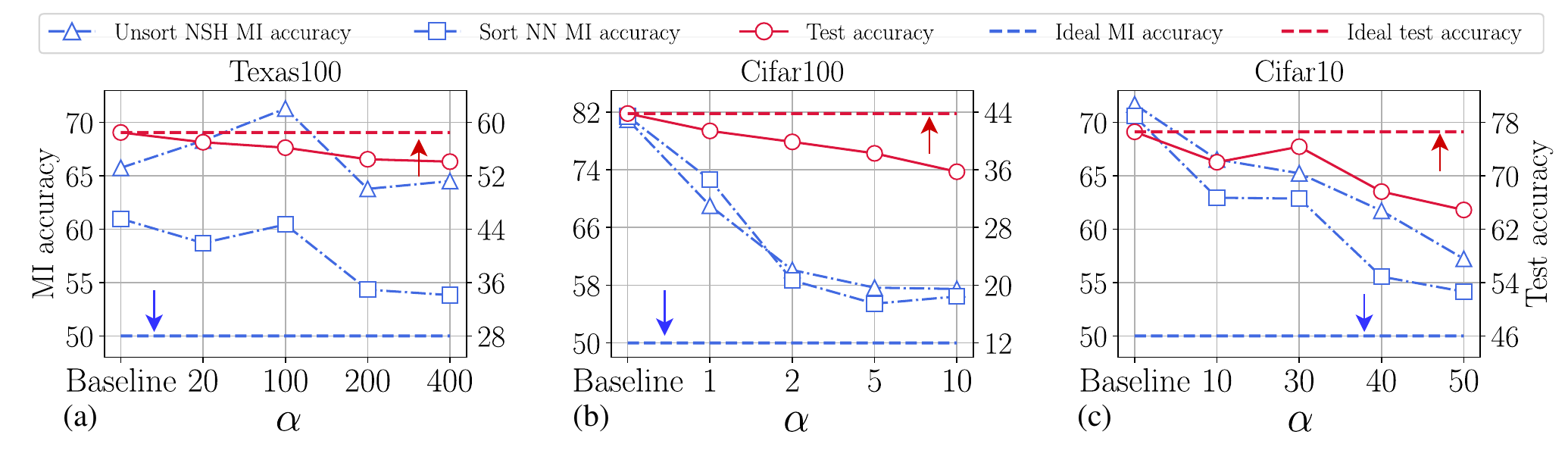}
\caption{\small Test accuracy and NN-based MI accuracy with different $\alpha$ for layer-wise balanced output control term. 
(a) Texas100, (b) CIFAR10, and (c) CIFAR100.
} 
\label{fig_balance}
\vspace{-10pt}
\end{figure}

\subsection{Hyperparameters setting and ablation study}

\subsubsection{Hyperparameter setting rules.}
\label{sec:hyperparametersetting}
Hyperparameter setting is essential to achieve effective defensive training by \textit{NeuGuard}. We provide a general guideline to tune the key parameters--$\alpha, \beta$ for layer-wise balanced output control and class-wise variance minimization.
1) for \textbf{$\beta$} which minimizes the variance, its initial value can be set as 10 times \#classes, e.g. $\beta=100$ (1000) for CIFAR10 (CIFAR100/Texas100). Then the maximum output confidence score $F(x)_{max}$ at each training batch can be obtained to guide the tuning: increase $\beta$ if $F(x)_{max}$ diverges rapidly from mean, and decrease $\beta$ if accuracy remains unchanged after first several epochs.
2) The initial values of $\alpha$ can be set to $\alpha=1,10,100$ for different models. We adopt the similar hyperparameter tuning approach that is used for $\beta$.
With above steps, the hyperparameters can easily enter into a suitable range that provides effective defense results with less sensitivity to their actual values. 
In the following, we investigate the individual contribution to defense of these two regularization terms through an ablation study.

\subsubsection{Effectiveness of class-wise variance minimization}

Fig.~\ref{fig_min_var_beta} shows the impact of $\beta$ on the testing accuracy and (both sorted and unsorted) MI accuracy obtained by our \textit{NeuGuard}, where only the class-wise variance minimization is used. We select $\beta$ roughly based on the number of classes in each dataset, and $\beta$ trades off the utility (i.e., testing accuracy) and privacy (i.e., MI accuracy). Specifically, when $\beta$ becomes larger, the class-wise variance should be smaller. This can enforce that all values in output confidence score vectors are closer, and thus reveal less information to the attacker, lowering the attack accuracy in general.  
On the other hand, similar values make output confidence score vectors not discriminative enough, which could negatively affect the testing accuracy. 
In summary, while the class-wise variance minimization provides an effective defense under certain conditions, that alone is insufficient for good generalizability against a range of attacks, e.g. far better defense against Unsorted NSH attack than Sorted NN attack.

\subsubsection{Effectiveness of the layer-wise balanced output control}
\label{balanced_output_control_ablation}
Fig.~\ref{fig_balance} (a), (b) and (c) show the test accuracy and NN-based MI accuracy when we only deploy the layer-wise balanced output control with different $\alpha$s for Texas100 dataset (using the fully connected neural network), CIFAR-100 and CIFAR-10 datasets (using CNNs), respectively. Here the convolution layers' output balancing also involves feature amplification, as discussed in Sec.~\ref{reg-method}. Its impact is discussed in detail in Appendix~\ref{eff_layer_amp}. 
We can observe that, unlike the results of only applying the variance minimization in Fig.~\ref{fig_min_var_beta}, applying layer-wise balanced output control generally can have more impact on the defense against the sorted NN attacks.
Particularly, it has stronger constraint on producing balanced intermediate results in each layer as $\alpha$ grows. This restricts the impact of some model parameters that may originally dominate the output prediction, in a layer-wise manner. 
Then the difference between the output predicted by the training data and testing data can be further reduced.

\subsubsection{Summary.}
The above analysis shows that our class-wise variance minimization contributes more on defending against the unsort NSH attack and layer-wise balanced output control can further reduce the sort NN attack. Therefore, \textit{NeuGuard} consisting of these two terms can better defend against both kinds of NN based MIAs as demonstrated in Table~\ref{tbl:nn_attack}.

\begin{table}[t]\renewcommand{\arraystretch}{}
\caption{\small Evaluation on metric based MI attacks.}
\vspace{-5pt}
\center
\footnotesize
\resizebox{\linewidth}{!}{%
\begin{tabular}{|c|c|c|c|c|c|c|}
\hline
\multicolumn{2}{|c|}{\textbf{Texas100}} & Baseline    & MemGuard    & Early stopping & AdvReg & \textbf{\textit{NeuGuard}}            \\ \hline\hline
\multicolumn{2}{|c|}{Testing accuracy}  & \textbf{56.0} & \textbf{56.0} & 54.5           & 54.0     & 54.4          \\ \hline
\multicolumn{2}{|c|}{Accuracy gap}      & 30.4        & 30.4        & 26.1           & 29.0     & \textbf{19.6} \\ \hline\hline
\multirow{4}{*}{\begin{tabular}[c]{@{}c@{}}MI\\ accuracy\end{tabular}} & Correctness & 65.2 & 65.2 & 63.1 & 64.5 & \textbf{59.8} \\ \cline{2-7} 
           & Confidence                 & 69.1        & 68.1        & 65.8           & 66.6   & \textbf{62.6} \\ \cline{2-7} 
           & Entropy                    & 62.2        & 60.6        & 60.5           & 60.6   & \textbf{54.4} \\ \cline{2-7} 
           & Modified entropy           & 69.1        & 68.4        & 65.7           & 66.8   & \textbf{62.7} \\ \hline
\end{tabular}
}
\center
\resizebox{\linewidth}{!}{%
\begin{tabular}{|c|c|c|c|c|c|c|}
\hline
\multicolumn{2}{|c|}{\textbf{CIFAR100}}                                                       & Baseline & MemGuard & Early stopping & AdvReg & \textbf{\textit{NeuGuard}}\\ \hline\hline
\multicolumn{2}{|c|}{Testing accuracy} & \textbf{43.8} & 42.9 & 41.8 & 39.7 & 43.0            \\ \hline
\multicolumn{2}{|c|}{Accuracy gap}     & 56.2          & 30.3 & 24.2 & 25 & \textbf{17.4} \\ \hline\hline
\multirow{4}{*}{\begin{tabular}[c]{@{}c@{}}MI\\ accuracy\end{tabular}} & Correctness & 78.1     & 65.1     & 61.9           & 62.5   & \textbf{58.7}             \\ \cline{2-7} 
           & Confidence                & 84.8          & 65.0   & 62.6 & 63.1 & \textbf{59.7} \\ \cline{2-7} 
           & Entropy                   & 80.8          & 56.1 & 59.6 & \textbf{55.5 }& 55.7 \\ \cline{2-7} 
           & Modified entropy          & 84.8          & 64.9 & 62.9 & 63.1 & \textbf{59.6} \\ \hline
\end{tabular}
}
\center
\resizebox{\linewidth}{!}{%
\begin{tabular}{|c|c|c|c|c|c|c|}
\hline
\multicolumn{2}{|c|}{\textbf{CIFAR10}}          & Baseline      & MemGuard      & Early stopping & AdvReg & \textbf{\textit{NeuGuard}}            \\ \hline\hline
\multicolumn{2}{|c|}{Testing accuracy} & \textbf{76.6} & \textbf{76.6} & 74.1           & 71.1   & 74.6          \\ \hline
\multicolumn{2}{|c|}{Accuracy gap}     & 23.4          & 23.4          & 22.2           & 21.0     & \textbf{17.8} \\ \hline\hline
\multirow{4}{*}{\begin{tabular}[c]{@{}c@{}}MI\\ accuracy\end{tabular}} & Correctness & 61.7 & 61.7 & 61.1 & 60.5 & \textbf{58.9} \\ \cline{2-7} 
           & Confidence                & 72.3          & 64.8          & 63.3           & 61.0     & \textbf{60.4} \\ \cline{2-7} 
           & Entropy                   & 70.5          & 58.6          & 60.4           & 58.0     & \textbf{55.9} \\ \cline{2-7} 
           & Modified entropy          & 72.1          & 65.0            & 63.2           & 61.1   & \textbf{60.4} \\ \hline
\end{tabular}
}
\label{tbl:metric_attack}
\vspace{-10pt}
\end{table}

\section{evaluation on metrics attacks}
\label{metric_base_evaluation}

In this section, we further evaluate our defense against the latest metrics based MI attacks~\cite{song2021systematic,choquette2021label}, as introduced in Sec.~\ref{intro_MIA}. 

\noindent \textbf{Experimental results comparison.}
Table~\ref{tbl:metric_attack} reports the testing accuracy, accuracy gap between training and testing set, attack accuracy based on the four metrics  (i.e., prediction correctness, prediction confidence, prediction entropy and modified prediction entropy) on CIFAR100, CIFAR10, and Texas100, respectively. For these attacks, \ul{we have the following observations:} 


\noindent \textbf{Our \textit{NeuGuard} always achieves the best results.} 
This conclusion is consistent with the results on NN based attacks, demonstrating the superior generality and scalability of our \textit{NeuGuard}.

\noindent \textbf{Our \textit{NeuGuard} has the smallest gap between training accuracy and testing accuracy}, which further implies that the output distribution of training set and that of testing set are the most similar. 
Here, we further propose to  verify this claim quantitatively. Specifically, we adopt three  metrics, i.e., 
Euclidean distance ($Euc$)~\cite{tabak2014geometry}, 
Kullback–Leibler divergence ($KL$)~\cite{mackay2003information},
Total Variation Distance  ($TV$)~\cite{levin2017markov}, 
 which are used to measure the similarity between two distributions---a smaller value of these metrics indicates a larger similarity of two distributions. 
 We denote the two probability distributions as $P$ and $Q$, respectively.
For $P$ and $Q$ in the same probability space $\Omega$, we have
$KL(P \| Q)=\sum_{\omega \in \Omega} P(\omega) \log \left(\frac{P(\omega)}{Q(\omega)}\right)$ and 
$TV(P, Q)=\frac{1}{2} \sum_{\omega \in \Omega}|P(\omega)-Q(\omega)| $.

 In our evaluation, we use histograms to calculate the probability distribution of the training set $P$ and that of the testing set $Q$ after the strongest modified entropy attack is applied (one can see that all defenses achieved the worst  performance against the modified entropy attack in Table~\ref{tbl:metric_attack}).  
Table~\ref{tab:kld_all} displays the results of the three metrics on the three datasets. 
All the results are obtained by selecting the best-performing model  for each defense. 
We can observe that our \textit{NeuGuard} has the lowest values evaluated by all the three metrics and significantly outperforms the compared defenses. 
These results indicate that our \textit{NeuGuard} can produce models with smallest difference between the training set and testing set outputs. Thus, our defense has the best effect on metric based MIAs.

\noindent \textbf{\textit{NeuGuard} delivers the best defense effectiveness against the strong C{\&}W label-only attack.}
Table~\ref{cw_label} compares the testing accuracy (utility), the prediction correctness attack accuracy (MI correction as a baseline), and C{\&}W label attack accuracy among different defenses. 
Again, \textit{NeuGuard} outperforms all other defenses while offering great utility (e.g. closer to baseline) against the C{\&}W label attack. The C{\&}W attack accuracy is even 3.6\% (CIFAR10) and 3.4\% (CIFAR100) lower than the correctness attack baseline. This is because \textit{NeuGuard} refines model parameters during the training and assures more evenly distributed output confidence scores.  
As a result, the needed adversarial perturbations to generate adversarial examples that alter prediction results to untargeted incorrect labels become similar for the training and testing data.

\begin{table}[t]\renewcommand{\arraystretch}{}
\center
\caption{\small Euclidean Distance, KL Divergence, and TV Distance of the  empirical output distributions.
}
\scriptsize
\addtolength{\tabcolsep}{-3pt}
\begin{tabular}{||c||ccc||ccc||ccc||}
\hline
            & \multicolumn{3}{c||}{{\bf Texas100}}   & \multicolumn{3}{c||}{{\bf CIFAR100}}    & \multicolumn{3}{c||}{{\bf CIFAR10}}    \\ \hline
Metrics     & Euc   & KL & TV    & Euc    & KL & TV    & Euc   & KL & TV    \\ \hline\hline
Baseline    & 92.88  & 0.2971        & 0.3265 & 132.44  & 0.8487        & 0.5829 & 324.04 & 0.3147        & 0.2700   \\ \hline
AdvReg      & 100.41 & 0.2441        & 0.2964 & 1127.68 & 0.3013        & 0.3092 & 76.97  & 0.1467        & 0.1864 \\ \hline
MemGuard    & 654.16 & 0.2645        & 0.2854 & 649.26  & 0.2896        & 0.3148 & 323.57 & 0.3229        & 0.2625 \\ \hline
Early stopping  & 72.21  & 0.2142        & 0.2717 & 104.19  & 0.1462        & 0.2212 & 182.77  & 0.2198       & 0.2435
 \\\hline
\textit{NeuGuard} & {\bf 7.39}   & {\bf 0.0943}        & {\bf 0.1636} & {\bf 11.93}   & {\bf 0.0681}        & {\bf 0.1578} & {\bf 6.68}   & {\bf 0.1151}        & {\bf 0.1840}  \\ \hline
\end{tabular}
\vspace{-5pt}
\label{tab:kld_all}
\end{table}

\noindent \textbf{Defenses against metrics based attacks are not as effective as against NN based attacks.} When comparing the results in Table~\ref{tbl:metric_attack} with those in Table~\ref{tbl:nn_attack}, we can see that metrics based attacks (except entropy) have larger MI accuracies than NN based attacks. This is because the performance of defenses against metrics based attacks, except entropy, are bounded by the accuracy gap between the training set and the testing set. When the accuracy gap is larger, all the defenses would achieve worse performance. 
The accuracy gap fully determine the prediction correctness attack, more detailed description and mathematical analysis can be found in Appendix~\ref{metric_defense_analysis}. 
In other words, \textbf{as long as the accuracy gap exists, no defense can reduce the  MI accuracy obtained by the prediction correctness attack to 0.5, i.e., random guessing.} 
The best defense performance we can achieve is to bring down the MI accuracy based on confidence prediction and modified prediction entropy attacks close to the correctness attack accuracy. 
We analyze two corner cases to further illustrate these metric-based attacks are bounded by the prediction correctness in Appendix~\ref{metric_defense_analysis}.

\begin{table}[!t]\renewcommand{\arraystretch}{}
\caption{\small C\&W label-only attack results on CIFAR datasets.}
\vspace{-5pt}
\resizebox{\linewidth}{!}{%
\begin{tabular}{||c|c||c|c|c|c|c||}
\hline
\textbf{Dataset} &
  \textbf{Accuracy} &
  \textbf{Baseline} &
  \multicolumn{1}{l|}{\textbf{MemGuard}} &
  \textbf{Early stopping} &
  \textbf{AdvReg} &
  \textbf{\textit{NeuGuard}} \\ \hline\hline
\multirow{3}{*}{\textbf{\begin{tabular}[c]{@{}c@{}}CIFAR\\ 10\end{tabular}}}  & Testing dataset   & \textbf{76.6} & \textbf{76.6} & 74   & 71.9          & {74.6} \\ \cline{2-7} 
& MI correctness    & 61.7          & 61.7          & 59.1 & \textbf{58.1} & 58.9          \\ \cline{2-7} 
& C\&W label attack & 69.2          & 69.2          &  59.7    & 59.2          & \textbf{55.3} \\ \hline\hline
\multirow{3}{*}{\textbf{\begin{tabular}[c]{@{}c@{}}CIFAR\\ 100\end{tabular}}} & Testing dataset   & \textbf{44.8} & \textbf{44.8} & 41.6 & 39.7          & {43}   \\ \cline{2-7} 
& MI correctness    & 77.5          & 77.5          & 61.6 & 64.6          & \textbf{57.8} \\ \cline{2-7} 
& C\&W label attack & 80.9          & 80.9          &  61.7    & 63.3          & \textbf{54.4} \\ \hline
\end{tabular}

\label{cw_label}
}
\vspace{-10pt}
\end{table}

\section{\textit{NeuGuard} against other MIAs}
\label{sect:recent_mia}
We also discuss and evaluate \textit{NeuGuard} against some newly proposed MIAs. 
For instance, \cite{yuan2022membership} proposed a self-attention MIA (SAMIA) focusing on enlarging the prediction KL divergence between training and testing data. Our \textit{NeuGuard} can well defend against SAMIA, as its objective is actually to minimize the prediction KL divergence---Table~\ref{tab:kld_all} shows that \textit{NeuGuard} significantly reduces the KL divergence. We also conduct experiments on CIFAR10/100 datasets by following the setting from~\cite{yuan2022membership}. Particularly, our \textit{NeuGuard} can reduce attack accuracy from 77.73\% and 69.29\% 
on Alexnet for CIFAR100 and CIFAR10 to 57.95\% and 51.37\%, respectively.
\cite{hui2021practical} proposed a blind membership inference
attack called BLINDMI, wherein the threat model only assumes knowledge of the output distribution, and nothing about the target model’s
architecture or training dataset. Our considered threat model 
is stronger, as part of the training/testing data is
available to the attacker, and the attacker also has the ability to query sufficient data samples from the target model. 
BLINDMI includes two different variants, i.e., BLINDMI-1CLASS and BLINDMI-DIFF. BLINDMI-1CLASS leverages the one-class SVM to learn the one-class semantics of nonmember labeled samples, 
while BLINDMI-DIFF iteratively performs differential comparison to infer membership. 
Table~\ref{tab:blindmi} shows our \textit{NeuGuard} defense results against them. We can observe that our \textit{NeuGuard} can effectively defend against BLINDMI-1CLASS and  BLINDMI-DIFF on both CIFAR10 and CIFAR100 datasets. 

In addition to attacking the traditional (non-defended) NN models for classification tasks, several recent works study MIAs 
against defended NN models (e.g., pruned sparse models~\cite{yuan2022membership}, adversarially trained models~\cite{song2019membership}), GANs for unsupervised learning~\cite{hayes2019logan, liu2019performing, chen2020gan}, federated learning~\cite{ melis2019exploiting,truex2019demystifying,he2020transnet}, etc. 
Directly 
integrating \textit{NeuGuard} into these models to defending against the MIAs is challenging because these methods have different goals as that of \textit{NeuGuard}.
However, our key idea of controlling inter-neuron and output is flexible and exploring this idea in the above models to better defend against MIAs datasets will be an interesting future work.

\section{Related work}
\label{related_work}

Many methods have been proposed to defend against membership inference attacks (MIAs). We categorize existing defenses as 
training time based defense (e.g.,   dropout ~\cite{salem2018ml}, $L_2$ norm regularization ~\cite{shokri2017membership}, model stacking~\cite{salem2018ml}, label smoothing~\cite{szegedy2016rethinking}, min-max adversary regularization (AdvReg)~\cite{nasr2018machine}, differential privacy~\cite{abadi2016deep,mohassel2017secureml}, early stopping~\cite{song2021systematic}, knowledge distillation~\cite{shejwalkar2021membership}) and inference time based defense (e.g., output perturbation (MemGuard)~\cite{jia2019memguard}).

\textbf{Label Smoothing (LS)}~\cite{szegedy2016rethinking} is a hard label augmentation approach to reduce model overfitting by assigning uniform probabilities to classes. However, studies~\cite{kaya2021does,muller2019does} show that LS pushes the model to output smoother and more uniform probabilities for the training data, and erases less information on testing data than training data. As a result, LS leads to greater discrepancies between the training data and testing data predictions, and thus increases model's MIA vulnerabilities. However, \textit{NeuGuard} exploits privacy-dedicated regularizations to constrain inner and output neurons with the goal of restricting the value range of output prediction during training. This explicitly reduces the prediction difference between training set and test set, and thus can defend against MIAs.

\label{sect:defenseDP}
\textbf{Differential privacy (DP)}~\cite{dwork2006calibrating} is a probabilistic privacy mechanism that provides an information-theoretical privacy guarantee. Many works aim to integrate DP into machine learning as a general approach to provide theoretical privacy guarantees for the models~\cite{abadi2016deep,rahman2018membership}. The basic idea is to add noise to the gradient used for the stochastic gradient descent~\cite{abadi2016deep,song2013stochastic,yu2019differentially} or the objective function for the model learning~\cite{iyengar2019towards} to achieve DP.
The main drawback of the current DP mechanism is that it cannot provide a satisfactory privacy-utility trade-off. As evaluated in ~\cite{jayaraman2019evaluating,rahman2018membership,shejwalkar2021membership} and discussed in ~\cite{jia2019memguard}, while DP demonstrates defense effectiveness against MIAs, its resulting model utility is fairly low. For example, ~\cite{shejwalkar2021membership} showed that DP-SGD~\cite{abadi2016deep} only delivers a $52.2\%$ testing accuracy for the Alexnet-CIFAR10 setting when the MIA accuracy is $51.7\%$, while \textit{NeuGuard} offers a $74.6\%$ testing accuracy at a similar level of membership privacy leakage.

\textbf{Knowledge distillation}~\cite{hinton2015distilling,komodakis2017paying} uses teacher-student models for model training. Distillation For Membership Privacy (DMP)~\cite{shejwalkar2021membership} firstly trains a teacher model with unprotected data and then trains the target model using an unlabeled reference dataset that aims to obtain a similar prediction entropy with the private training data. With limited access to the private data, the trained student model should not leak information.
Compared to our \textit{NeuGuard}, this design requires more complex training procedures and additional reference datasets to facilitate the knowledge distillation between the teacher model and student model.
However, public reference datasets from the same domain may not be available in real-world applications, especially in fields like medical and financial analysis, where the data are private and confidential. 
Furthermore, one alternative solution to this is to generate synthetic data using the generative adversarial networks (GANs). However, ~\cite{shejwalkar2021membership} shows that it would result in significant model utility drop. 
~\cite{chourasia2022knowledge}, ~\cite{tang2021mitigating} improve the defense without requiring public data, but they require more complex training strategies, different training phases for teacher and student models, and require sufficient training data since they need to split data into several parts for training.

\begin{table}[!t]\renewcommand{\arraystretch}{}
\caption{\small \textit{NeuGuard} against BLINDMI on CIFAR datasets.}
\vspace{-5pt}
\label{tab:blindmi}
\resizebox{\linewidth}{!}{%
\begin{tabular}{|c|cccc|cccc|}
\hline
\textbf{Attack} &
  \multicolumn{4}{c|}{\textbf{BLINDMI-1CLASS}} &
  \multicolumn{4}{c|}{\textbf{BLINDMI-DIFF}} \\ \hline
Dataset &
  \multicolumn{2}{c|}{CIFAR100} &
  \multicolumn{2}{c|}{CIFAR10} &
  \multicolumn{2}{c|}{CIFAR100} &
  \multicolumn{2}{c|}{CIFAR10} \\ \hline
Model &
  \multicolumn{1}{c|}{Baseline} &
  \multicolumn{1}{c|}{\textbf{\textit{NeuGuard}}} &
  \multicolumn{1}{c|}{Baseline} &
  \textbf{\textit{NeuGuard}} &
  \multicolumn{1}{c|}{Baseline} &
  \multicolumn{1}{c|}{\textbf{\textit{NeuGuard}}} &
  \multicolumn{1}{c|}{Baseline} &
  \textbf{\textit{NeuGuard}} \\ \hline
Test Acc &
  \multicolumn{1}{c|}{41.36} &
  \multicolumn{1}{c|}{\textbf{42.93}} &
  \multicolumn{1}{c|}{77.83} &
  \textbf{74.47} &
  \multicolumn{1}{c|}{41.36} &
  \multicolumn{1}{c|}{\textbf{42.77}} &
  \multicolumn{1}{c|}{77.85} &
  \textbf{74.47} \\ \hline
Attack Acc &
  \multicolumn{1}{c|}{60.10} &
  \multicolumn{1}{c|}{\textbf{50.00}} &
  \multicolumn{1}{c|}{55.48} &
  \textbf{50.00} &
  \multicolumn{1}{c|}{71.15} &
  \multicolumn{1}{c|}{\textbf{55.24}} &
  \multicolumn{1}{c|}{68.01} &
  \textbf{56.65} \\ \hline
Precision &
  \multicolumn{1}{c|}{65.18} &
  \multicolumn{1}{c|}{\textbf{0.00}} &
  \multicolumn{1}{c|}{57.32} &
  \textbf{0.00} &
  \multicolumn{1}{c|}{71.66} &
  \multicolumn{1}{c|}{\textbf{56.97}} &
  \multicolumn{1}{c|}{61.02} &
  \textbf{56.10} \\ \hline
Recall &
  \multicolumn{1}{c|}{43.37} &
  \multicolumn{1}{c|}{\textbf{0.00}} &
  \multicolumn{1}{c|}{42.94} &
  \textbf{0.00} &
  \multicolumn{1}{c|}{69.96} &
  \multicolumn{1}{c|}{\textbf{42.78}} &
  \multicolumn{1}{c|}{99.73} &
  \textbf{61.23} \\ \hline
F1 score &
  \multicolumn{1}{c|}{52.08} &
  \multicolumn{1}{c|}{\textbf{0.00}} &
  \multicolumn{1}{c|}{49.10} &
  \textbf{0.00} &
  \multicolumn{1}{c|}{70.80} &
  \multicolumn{1}{c|}{\textbf{48.87}} &
  \multicolumn{1}{c|}{75.71} &
  \textbf{58.55} \\ \hline
\end{tabular}
}
\vspace{-15pt}
\end{table}

\section{Conclusion}
\label{conclusion}

We explored state-of-the-art defenses and showed they are ineffective in defending against existing MIAs, especially sorted and unsorted NN attacks.
We advocate a more effective defense
is to orchestrate the output of the training set and testing set for the same explicitly designed distribution that is more evenly distributed in a restricted small range.
To achieve this goal, we propose a simple yet effective defense mechanism--\textit{NeuGuard}--built upon the technique of fine-grained neuron-level regularization, to simultaneously control and guide the final output neurons and hidden neurons towards constructing a defensive model.
\textit{NeuGuard} consists of a class-wise variance minimization  and layer-wise balanced output control to regularize output and inner neurons in a layer-wise manner.  
We validate the effectiveness of \textit{NeuGuard} on three different datasets against not only two NN based MIAs, but also five (strongest) metrics based MIAs including the label-only attack. 
We further discuss the defense upper bound of the metrics based MIAs through theoretical and experimental analysis.
With a flexible parameter control, \textit{NeuGuard} always offers the best utility-privacy trade-off with much lower overhead, comparing with all evaluated defenses.


\begin{acks}
We would like to thank the shepherd and the anonymous reviewers for their constructive comments and suggestions on this work. This work is partially supported by the National Science Foundation (NSF) under Award CCF-2011236, and Award CCF-2006748.
\end{acks}
\bibliographystyle{ACM-Reference-Format}
\bibliography{ref}

\appendix
\appendix
\section{Appendices}
\begin{table}[!h]
\caption{Model accuracy and NN based MI accuracy with different $\lambda$ values for the adversary regularization on Texas100.}
\vspace{-10pt}
\label{advreg_texas100}
\centering
\resizebox{0.8\linewidth}{!}{
\begin{tabular}{|c|c|c|c|c|c|}
\hline
\multicolumn{2}{|c|}{$\lambda$}        & {1} & {2} & \textbf{3} & \textbf{5} \\ \hline \hline
\multicolumn{2}{|c|}{{Training set accuracy}}  & 85.99      & 86.08      & 66.67      & 47.77      \\ \hline
\multicolumn{2}{|c|}{{Testing set accuracy}}  & 58.51      & 58.17      & 50.92      & 40.59      \\ \hline
\multirow{2}{*}{{\begin{tabular}[c]{@{}c@{}}MI \\ accuracy\end{tabular}}} & {Sorted NN}    & 68.56      & 68.31      & 64.18      & 55.81      \\ \cline{2-6} 
& {Unsorted NSH}     & 60.41      & 60.44      & 53.48      & 52.24      \\ \hline
\end{tabular}
}
\vspace{-15pt}
\end{table}

\subsection{Datasets}
\label{datasets}

We use the following three benchmark datasets  to demonstrate the effectiveness of our \textit{NeuGuard} against membership inference attack for different application scenarios.

\textbf{Texas100}~\cite{texas100} is a dataset generated from Hospital Discharge Data Public Use Data File that contains information about inpatients stays in several health facilities. The data is published by the Texas Department of State Health Services (DSHS) and we obtained the preprocessed dataset from \cite{shokri2017membership}. This dataset contains 67,330 data records with 6,170 binary features. These features indicate the external causes of injury (e.g., drug misuse and suicide), the diagnosis (e.g., schizophrenia, illegal abortion), the procedures the patient underwent (e.g., surgery), and generic information such as gender, age, race, hospital ID, and length of stay. 
The records are clustered into 100 classes representing the 100 most frequent medical procedures.
Following the existing methods~\cite{nasr2018machine, jia2019memguard, nasr2019comprehensive}, we use 10,000 data records for training and 57,330 data records for testing. 

\textbf{CIFAR10} and \textbf{CIFAR100} are benchmark datasets widely used in image classification tasks~\cite{krizhevsky2009learning}. Specifically, CIFAR-10 consists of 32$\times$32 color images from 10 classes and each class contains 6,000 images. It includes 50,000 training images and 10,000 testing images. CIFAR100 contains the same size color images from 100 non-overlapping classes, each with 500 training images and 100 testing images.
Table~\ref{data_size} summarizes the statistics of these datasets.

\begin{table}[!t]
\caption{Dataset split configurations.}
\vspace{-10pt}
\center
\resizebox{\linewidth}{!}{%
\begin{tabular}{|c|c|c|c|c|}
\hline
Dataset  & Training set & Testing set  & Training members & Training non-members \\ \hline\hline
CIFAR10  & 50,000 & 10,000 & 25,000           & 5,000                \\ \hline
CIFAR100 & 50,000 & 10,000 & 25,000           & 5,000                \\ \hline
Texas100 & 10,000 & 57,330 & 5,000            & 10,000               \\ \hline
\end{tabular}
}
\label{data_size}
\vspace{-2mm}
\end{table}


\subsection{Parameter setting}
\label{params}
For the {Texas100} classification task, we use a fully connected neural network with four hidden layers, which have layer sizes 1024, 512, 256, 128, respectively. We use the Tanh activation function for the hidden layers and use the softmax function for the final layer. We use the cross-entropy loss function and Adam optimizer to train the model. 
We train the model with an initial learning rate 0.001 and a decay by 0.1 in every 20 epochs. 

For CIFAR10 and CIFAR100 image classification tasks, we use Alexnet\cite{krizhevsky2012imagenet}, which is a convolutional neural network with parameters trained with a cross-entropy loss function and stochastic gradient descent (SGD) optimizer. 
We set the initial learning rate as 0.01 and decay by 0.1 in every 20 epochs. 

We follow the training method and hyperparameter setup proposed by the authors and use the published code to evaluate their 
defense mechanism.
The adversarial regularization parameter is set as $\lambda=3$ for Texas100 model and $\lambda=6$ for CIFAR10 and CIFAR100. 
The Early stopping model is introduced following the \cite{song2021systematic} to show the defense efficiency of the regular trained model with the similar test accuracy of the AdvReg trained model. 
We evaluate the MemGuard defense by using Sorted NN model~\cite{jia2019memguard} and Unsorted NSH model~\cite{nasr2018machine} as its defense classifiers to generate the noisy output, respectively. Since in some cases, the defense classifier with Sorted NN attack model under the constraint of MemGuard, is unable to produce effective noisy output for defense, 
we select the Unsorted NSH attack model which offers better defense effectiveness as the classifier, for fairly comparing the defense efficiency. 

{
\begin{algorithm}[!t]
\footnotesize
  \begin{algorithmic}[1]
    \STATE \textbf{Input: } ML model $F$, a batch of data $(x_{B},y_B)$ with $N$ records, class-wise mean list vector $\mu_{y}$, model layer number $M$
    \STATE \textbf{Output: } Loss value $Loss$ calculated for this batch
    \STATE $ \{outputs, h^1, h^2, ..., h^{M-1} \}  = F(x_B)$
    \STATE $softout = softmax(outputs)$
    \FOR {$l$ in $M-1$}
        \STATE {$L_{boc} =  L_{boc} + \frac{1}{S_{l}}\left\|\sum_{i=1}^{\left\lfloor S_{l} / 2\right\rfloor} h_{i}^{l}-\sum_{i=\left\lfloor S_{l} / 2\right\rfloor+1}^{S_{l}} h_{ i}^{l}\right\|_{F}^{2}$}
    \ENDFOR
    \FOR {$i$ in $N$}
        \STATE {$count_{y_i}  += 1$}
        \STATE $\mu_{y_i} = \mu_{y_i}\frac{count_{y_i}-1}{count_{y_i}} +\frac{softout_i}{count_{y_i}}$
    \ENDFOR
    \STATE {$L_{var} = \frac{1}{n} \sum_{i=0}^{n} (F(x_i)-\mu_y )^{2} $}
    \STATE $Loss = criterion(x_{B},y_B) +  \alpha \times L_{boc} + \beta \times L_{var} $
  \end{algorithmic}
  \caption{Loss calculation using proposed method}
  \label{alg_1}
\end{algorithm}
}

\begin{figure}[!t]
\centering
\includegraphics[width=\linewidth]{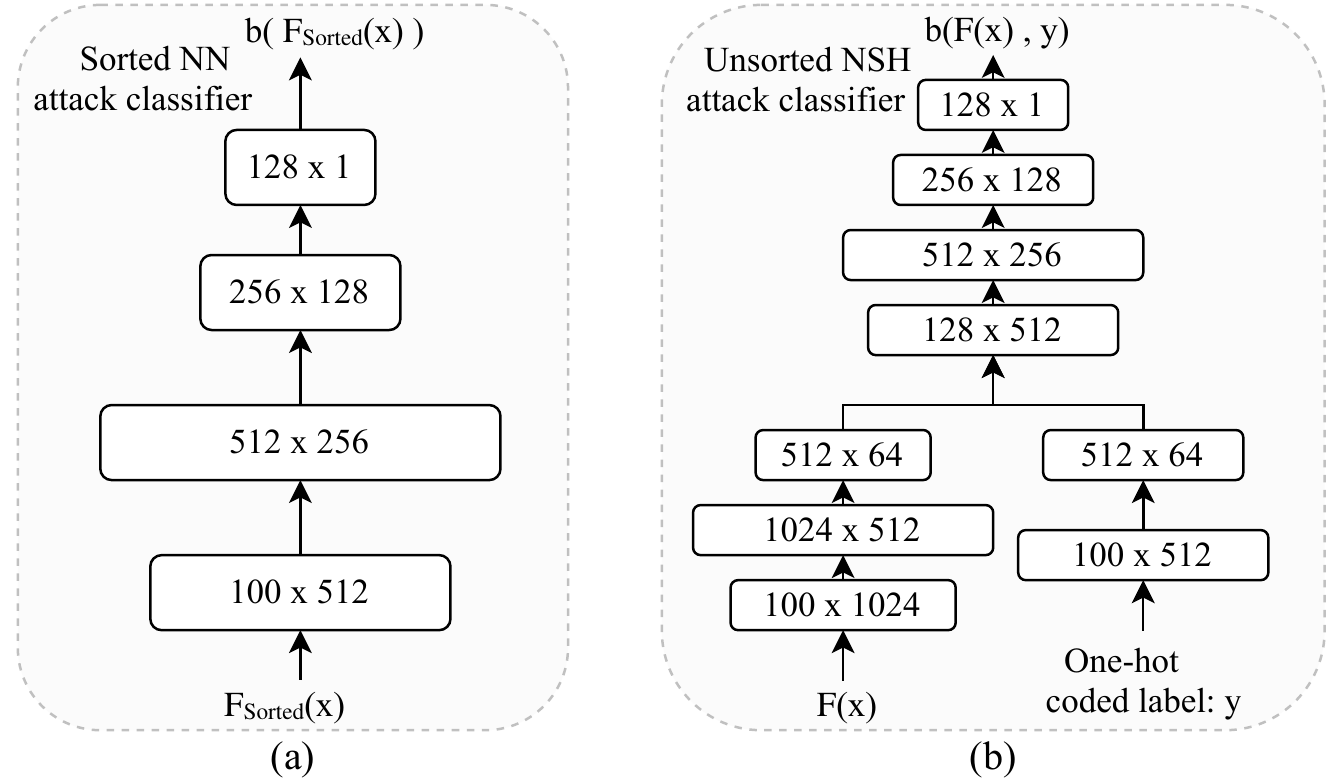}
\caption{The attack classifier architecture of Sorted NN and Unsorted NSH attack used in the evaluation }
\label{fig_attacknn_arch}
\vspace{-4mm}
\end{figure}


\subsection{Neural network based attacks setup}
\label{nn_attack_setup}
We summarize neural network based attacks as \emph{sorted attack} and \emph{unsorted} attack. 
The major difference is whether the output confidence score used by the the attack model is sorted or not.

\noindent \textbf{Sorted attack:}
For the Sorted NN attack, we adopt the standard setting in \cite{jia2019memguard} and use a three-layer fully connected neural network 
as the attack classifier. See  Fig.~\ref{fig_attacknn_arch}(a). 

\noindent \textbf{Unsorted attack:}
To evaluate the Unsorted NSH attack, we follow the model structure and setup in \cite{nasr2018machine} to construct and train the attack classifier. The Unsorted NSH attack classifier takes two pieces of information as input. One is the unsorted confidence score vector, and the other one is the one-hot encoded label (all elements except the one that corresponds to the label index are 0).
The classifier consists of three fully connected sub-networks. 
See Fig.~\ref{fig_attacknn_arch}(b). 

We adopt the Relu activation function for the hidden layers and the sigmoid activation function for the output layer for both attack classifiers. The attack classifier predicts the input as a member if and only if the final prediction probability $b(\cdot) \ge 0.5$; otherwise, it predicts as a non-member.
To train the attack classifier, we use the mean squared error (MSE) criterion and Adam optimizer with a learning rate of 0.001. For better convergence, we decay the learning rate by 0.1 in the 40th and 90th epoch for 100 epoch training for the Sorted NN attack, and decay the learning rate by 0.1 in the 30th epoch for 200 training epochs for Unsorted NSH attack.

\begin{figure*}[!tb]
\centering
\includegraphics[width=0.8\textwidth]{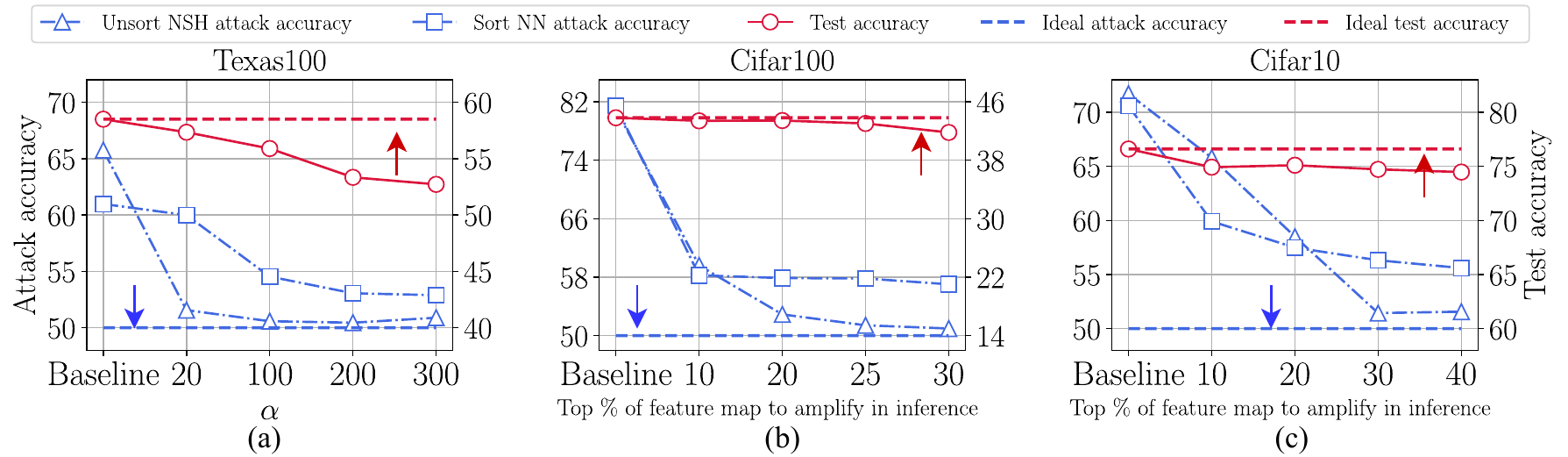}
\caption{Testing accuracy and NN-based MI  accuracy with different hyperparameters for layer-wise operations. For the fully connect model for Texas100 classification, we set $\beta=3,000$ in the class-wise variance minimization regularization, and we change the $\alpha$ value in (a) to show the effects of the layer-wise balanced output control. For convolutional models, we set $\beta=200$ on CIFAR10  and $\beta=1,000$ on CIFAR100. We  amplify the top $10\%$ feature map values for 1.5 and 2 times, respectively. (b) and (c) show the results with different selected top feature map values during the inference on CIFAR100 and  CIFAR10.} 
\label{fig_comb_tune}
\vspace{-3mm}
\end{figure*}


\begin{table}[!t]
\caption{Variance of the output confidence scores on the training set and testing set for CIFAR100.}
\label{mean_var_cifar100}
\vspace{-15pt}
\center
\resizebox{\linewidth}{!}{%
\begin{tabular}{|c|c|c|c|c|c|}
\hline
Model     & Baseline & Early stopping & AdvReg  & \begin{tabular}[c]{@{}c@{}}MemGuard\end{tabular} & \textit{NeuGuard} \\ \hline\hline
\begin{tabular}[c]{@{}c@{}}Training set\end{tabular} & 5.37E-03  & 4.57E-03    & 6.99E-03 & 4.30E-03                                                & \textbf{4.44E-06}    \\ \hline
\begin{tabular}[c]{@{}c@{}}Testing set\end{tabular}  & 4.08E-03  & 3.48E-03    & 5.89E-03 & 3.52E-03                                                & \textbf{3.88E-06 }   \\ \hline
\end{tabular}
}
\vspace{-16pt}
\end{table}

\subsection{Effectiveness of the layer-wise feature map operations}
\label{eff_layer_amp}

In this section, we show some supplementary results with combined \textit{NeuGuard} training methods and tune one of the component's hyperparameter to further demonstrate the effectiveness of the proposed methods.


Fig.~\ref{fig_comb_tune} (a) shows the test accuracy and NN-based MI accuracy when we vary $\alpha$ for layer-wise balanced output control on the Texas100 dataset. Here we set $\beta=3000$ for the class-wise variance minimization. Unlike the results obtained by only applying the variance minimization in Fig.~\ref{fig_min_var_beta} (a), with $\alpha$ growing, the attack accuracies of both Unsorted NSH attack and Sorted NN attack decrease prominently at the cost of limited utility reduction.
A larger $\alpha$ indicates a better defense effectiveness but a slightly degraded testing accuracy.  

As discussed in Sec.~\ref{reg-method} and Sec.~\ref{balanced_output_control_ablation}, we adopt the layer-wise feature map amplification for the convolutional layers in CNNs at both training and inference stages to 1) better assist the proposed learning regularization for converging model output scores towards desired distributions (for defense effectiveness); 2) better maintain model inference accuracy (for model utility).
For the hyperparameters setting of feature amplification, we can empirically amplify top $0-10\%$ feature maps to $1-1.5$ times during training guided by the accuracy growth trend. Then at inference we amplify top $0-50\%$ feature maps using the same amplification rate to reduce the accuracy gap with an acceptable utility loss.
To validate its effectiveness, we conduct experiments based on the CNN-based CIFAR100 and CIFAR10 image classification tasks and tune the amplified percentage of feature maps during the inference time.   
As Fig.~\ref{fig_comb_tune} (b) and (c) show,
the test accuracy on CIFAR100 and CIFAR10 drops slightly as the percentage of amplified features increases, while the attack accuracy of both Sorted NN and Unsorted NSH attack decreases more significantly, demonstrating a much improved trade-off between defense effectiveness and model utility using our method. This is because: 
1) 
The impact of the most significant portions of a feature map that are amplified during training decreases as other parts of the feature map is also enlarged due to the increased amplified percentage during inference; 2) The operation is 
analogous to  
introducing some noise to model inference, making the output distribution of training set and test set closer and resulting in reduced MI accuracy for both attacks. However, amplifying too many intermediate features will cause more obvious utility drop despite the minor attack accuracy reduction. Therefore, the optimal parameter of our proposed defense can be identified by observing MI accuracy reduction and utility loss, e.g. 25\% and 30\% for CIFAR100 and CIFAR10 as shown in Fig.~\ref{fig_comb_tune} (b) and (c).


\subsection{Defense effectiveness analysis for metric based attack}
\label{metric_defense_analysis}

We first show the close relationship between prediction correctness and accuracy gap between the training set and testing set. Then, we discuss why the defenses cannot reduce the MI accuracy of metrics based attacks to random guessing.

\begin{figure*}[!t]
\centering
\subfigure[Normal model]
{\includegraphics[width=0.24\textwidth]{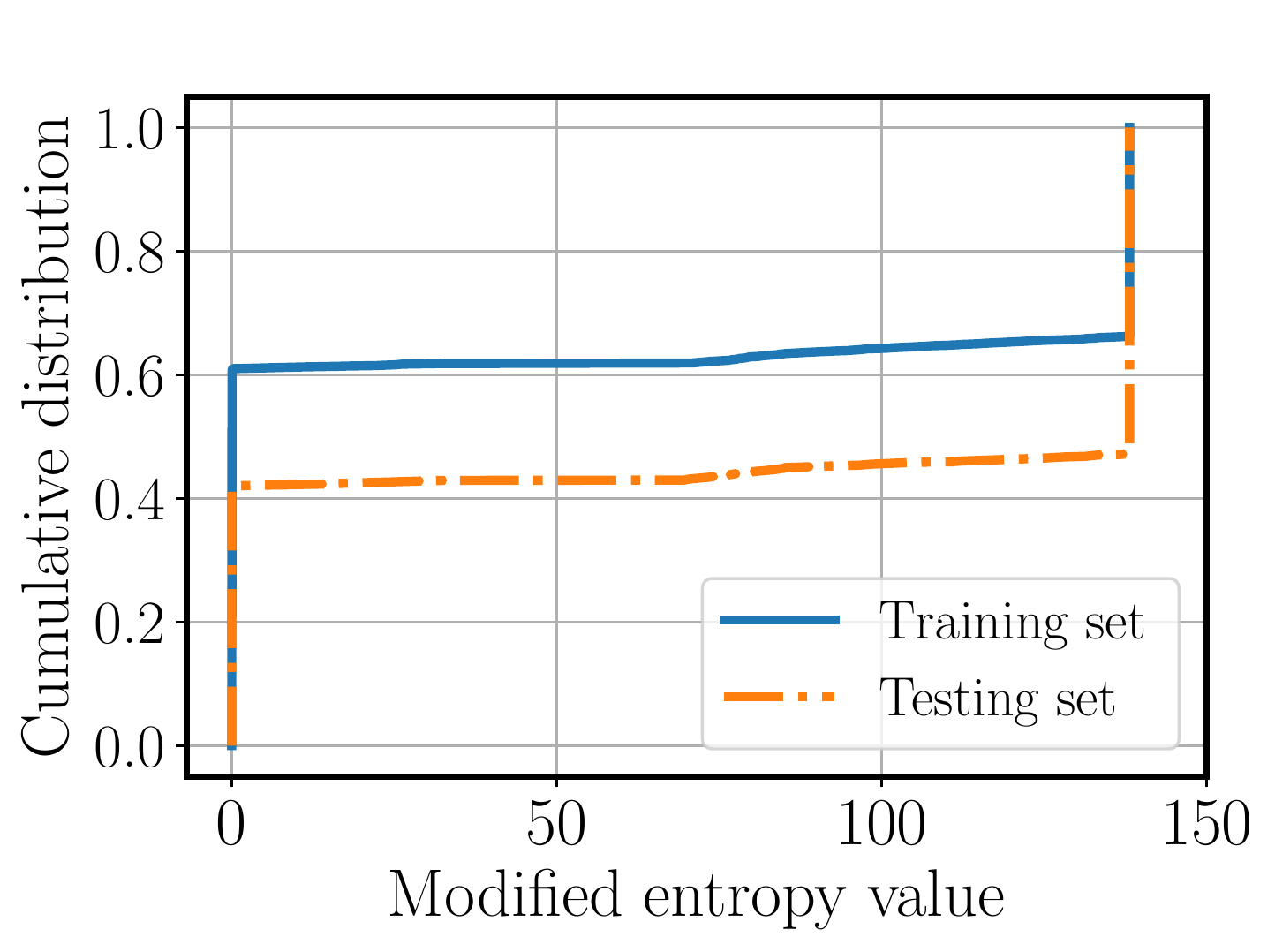} \vspace{-5pt}
\label{normal:fc}}
\subfigure[\textit{NeuGuard}]
{\includegraphics[width=0.24\textwidth]{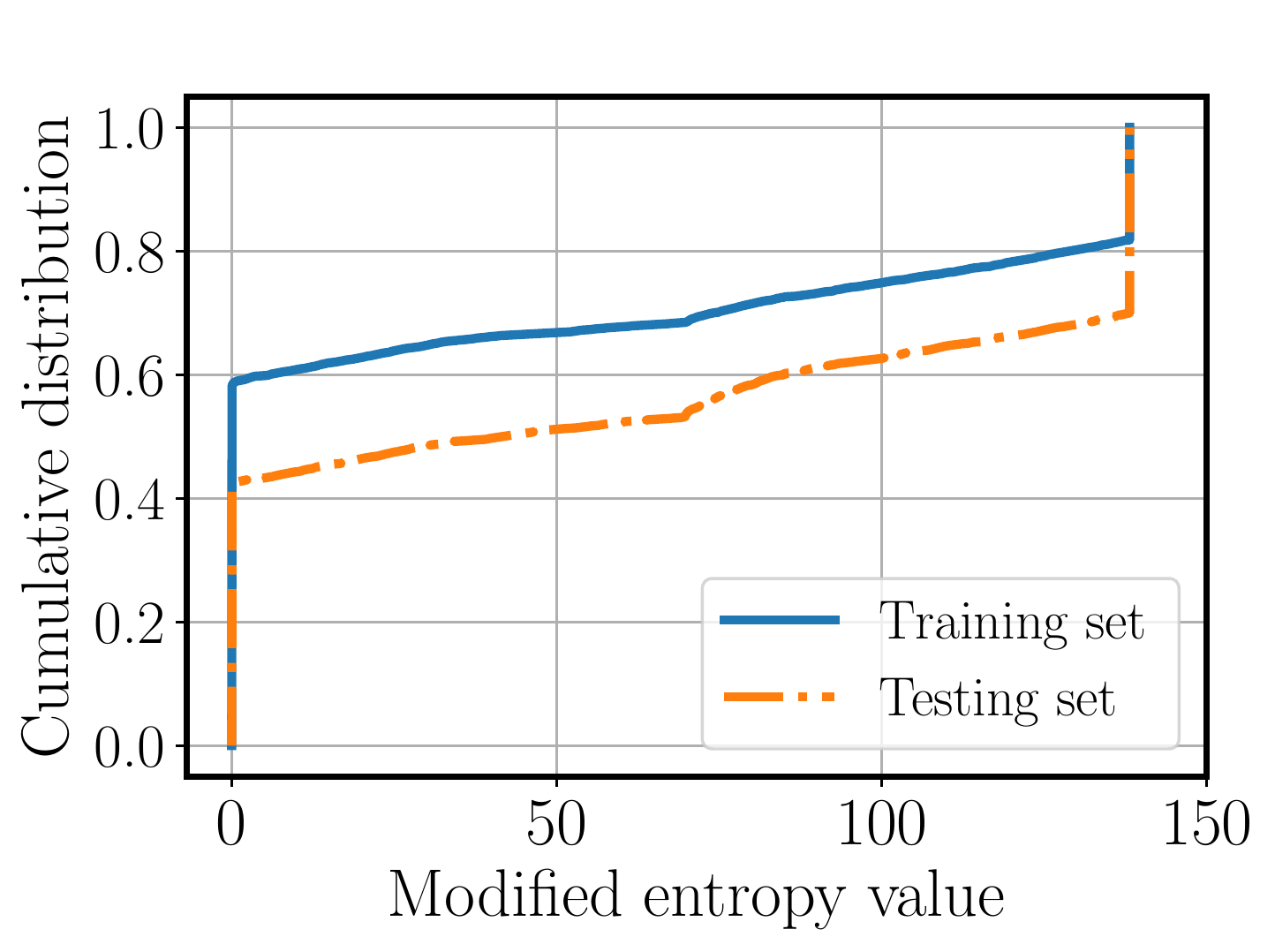} \vspace{-5pt}
\label{ourmodel:fc}}
\subfigure[Normal model]
{\includegraphics[width=0.24\textwidth]{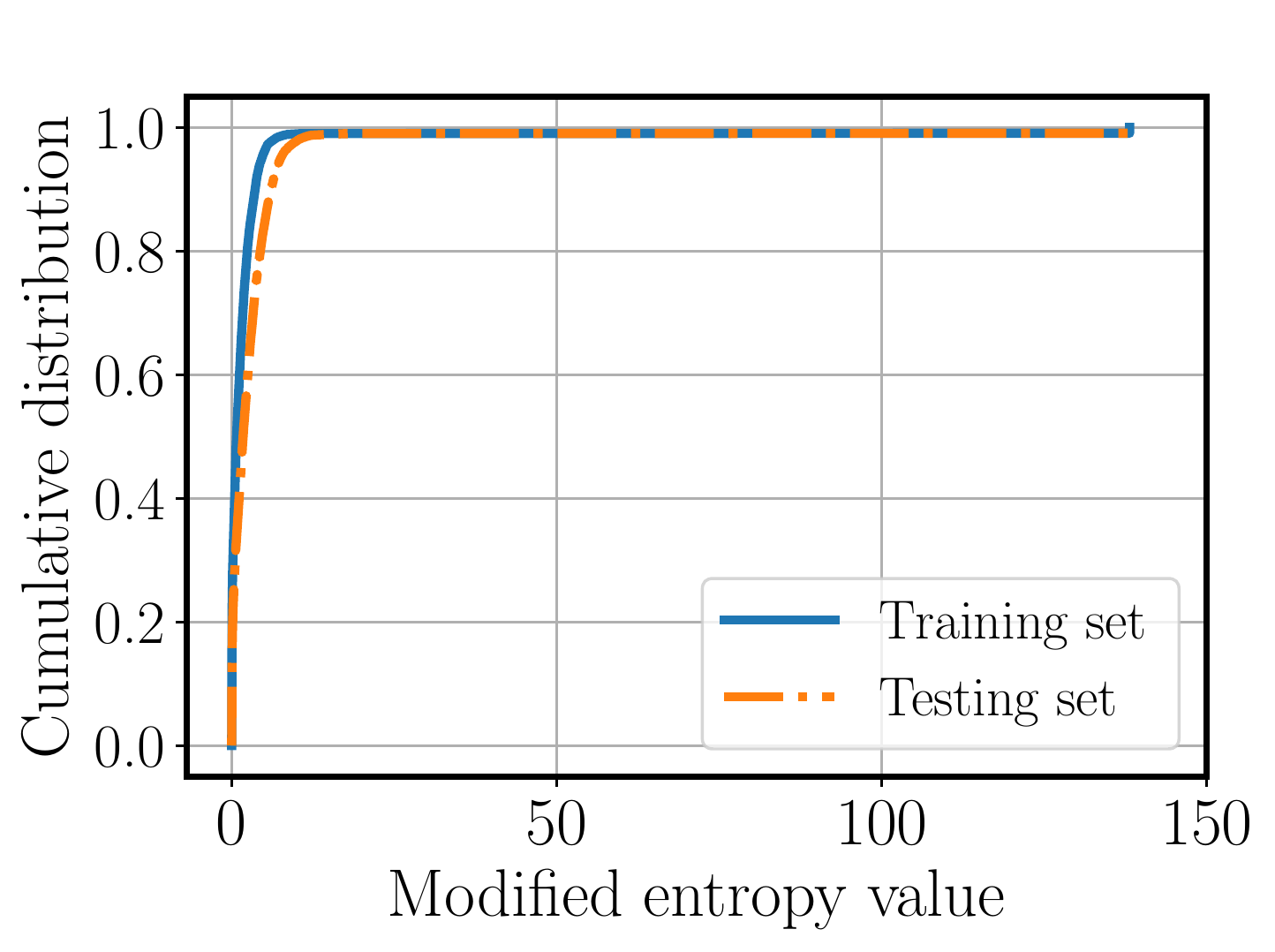} \vspace{-5pt}
\label{normal:conv}}
\subfigure[\textit{NeuGuard}]
{\includegraphics[width=0.24\textwidth]{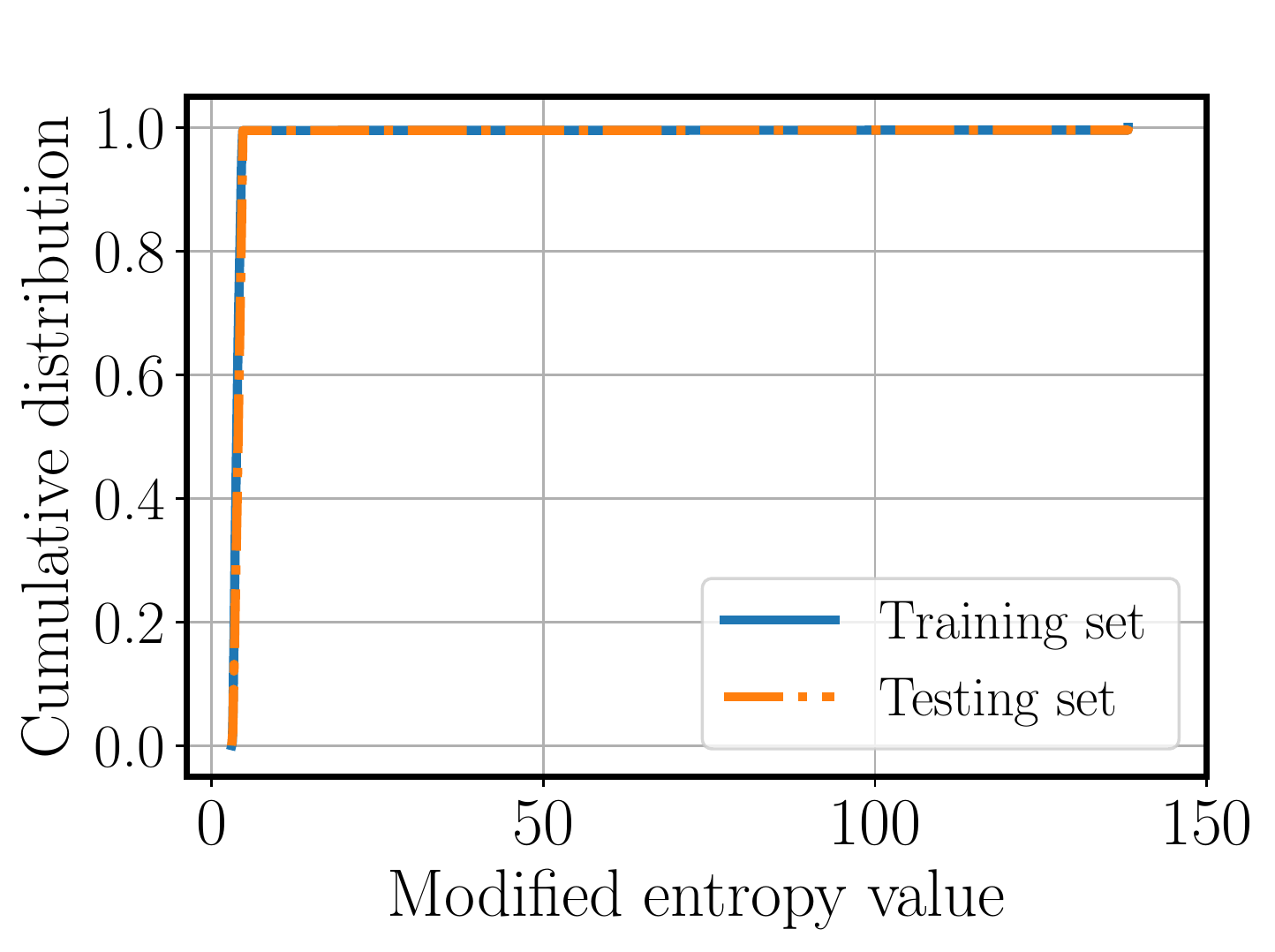}
\label{ourmodel:conv}}
\vspace{-5mm}
\caption{Empirical CDF of modified entropy on CIFAR100  with (a) and (b) fully connected layer amplification in inference. Amplify top 50\% for 20$\times$; and with (c) and (d) convolution layer amplification in inference. Amplify the top 1\% for 1000$\times$.}
\vspace{-8pt}
\end{figure*}

Let $d_{tr}$ and $d_{te}$ denote the number of training samples and testing samples, respectively. Note that in our experiments, we have the same value of $d_{tr}$ and $d_{te}$ and we let $d_{tr} = d_{te} = d$. Moreover, we denote the training accuracy and testing accuracy as $Acc_{tr}$ and $Acc_{te}$, respectively. Then, we have:  
\begin{footnotesize}
\vspace{-2pt}
\begin{equation}
\begin{aligned}
 \label{pre_correct}
\mathcal{M}_{\text {corr }}(F ; z)&=I(\operatorname{argmax} F(x)=y) 
= \frac{Acc_{tr}\times d_{tr}+(d_{te}-Acc_{te}\times d_{te})}{d_{tr}+d_{te}}  \\
&= \frac{d \times(Acc_{tr}-Acc_{te})+d}{2d}  
= \frac{1}{2}\times Acc_{gap} + \frac{1}{2} 
\end{aligned}
\end{equation}
\end{footnotesize}
Eqn~\ref{pre_correct} implies that the prediction correctness can be completely determined by the accuracy gap. In other words, \textbf{as long as the accuracy gap exists, no defense can reduce the  MI accuracy obtained by the prediction correctness attack to 0.5, i.e., random guessing.} 
Furthermore, from Eqn~\ref{eqn:pc} and Eqn~\ref{eqn:mpe}, we note that \textbf{prediction confidence and modified prediction entropy are also highly related to prediction correctness, as they both use the predicted label information.} In an extreme case where the model outputs the same distribution of confidence score vectors, e.g. a single large score with all others being equally small values, regardless of member or non-member data, the calculation of both prediction confidence and modified prediction entropy can converge to two values--one value for all correctly predicted data and another for all incorrectly predicted data, e.g. CDF with only two values in the x-axis as we shall show in Fig.~\ref{normal:fc}. As a result, their  attack accuracies are the same as that of prediction correctness for any preset threshold chosen between these two values.    

\textbf{The best defense performance we can achieve is to bring down the MI accuracy based on confidence prediction and modified prediction entropy attacks close to the correctness attack accuracy. }
We explore the two cases on CIFAR100  through the proposed layer-wise intermediate results amplification to demonstrate that the defense performance is upper bounded by applying the prediction correctness attack. 
We need to point out that this kind of defense is easy to achieve through our layer-wise amplification. In this case, 
these metric based attacks are equivalent to the prediction correctness attack.

Table~\ref{cifar100_corner_case} shows the testing accuracy and metric based attack results on two cases that lead the attack accuracy of prediction confidence and  modified prediction entropy close to the prediction correctness attack. We apply two amplification strategies to the model and both of them can maintain the test accuracy at the same level as the baseline model. 
These two strategies show the analysis of two extreme cases. The choice of hyperparameters is intended to amplify the amplification impact of certain feature maps while maintaining the accuracy of the model.
The first strategy is 
to amplify top 50\% of the intermediate results of the last three FC layers for $20\times$. Fig.~\ref{normal:fc} and Fig.~\ref{ourmodel:fc} show the absolute modified entropy distribution of this strategy applied on the regular trained model and \textit{NeuGuard}. The output confidence score is amplified to the case that only one label has large value  (e.g., close to 1) and all the others have equally small values (e.g., close to 0). In this case, the modified prediction entropy for the correctly classified data is 0 and for the misclassified data is about 138. As the figure shows, the only difference in the modified entropy value is made by the correctness of the model classification.
The attack accuracies of prediction correctness, confidence and modified entropy are similar as shown in Table~\ref{cifar100_corner_case} for both the regular trained model and \textit{NeuGuard}.

\begin{table}[!t]\renewcommand{\arraystretch}{0.9}
\caption{Metric based eval. on two corner cases on CIFAR100.}
\vspace{-10pt}
\center
\resizebox{\linewidth}{!}{%
\begin{tabular}{||c||c|c|c|c||}
\hline
Model &
  \begin{tabular}[c]{@{}c@{}}Baseline\\ FC layer amp\end{tabular} &
  \begin{tabular}[c]{@{}c@{}}\textit{NeuGuard}\\ FC layer amp\end{tabular} &
  \begin{tabular}[c]{@{}c@{}}Baseline\\ Conv layer amp\end{tabular} &
  \begin{tabular}[c]{@{}c@{}}\textit{NeuGuard}\\ Conv layer amp\end{tabular} \\ \hline\hline
Test accuracy      & 43.6 & 43.4 & 43   & 43   \\ \hline
Accuracy gap       & 26.6 & 17.1 & 25.6 & 17.3 \\ \hline\hline
Correctness        & 63.2 & 58.5 & 62.8 & 58.7 \\ \hline
Confidence         & 63.7 & 58.9 & 63.5 & 59.7 \\ \hline
Entropy            & 51.7 & 55.6 & 55.5 & 55.4 \\ \hline
Modified   entropy & 63.2 & 58.9 & 63.8 & 59.7 \\ \hline
\end{tabular}
}
\label{cifar100_corner_case}
\vspace{-5mm}
\end{table}

The second case is explored by amplifying the most significant part of the feature map in the convolution layers while ensuring the model utility. In the experiments, we amplify the top 1\%  feature map values to $1000\times$ for all the convolution layers. The absolute modified entropy distribution of the normal model and \textit{NeuGuard} are shown in Fig.~\ref{normal:conv} and Fig.~\ref{ourmodel:conv} and the results are in Table~\ref{cifar100_corner_case}. The modified prediction entropy has almost no difference for training set and testing set as all of them are close to zero. Nevertheless, the attack accuracy of the modified prediction entropy is still similar to that of the prediction correctness in both the regular trained model and our \textit{NeuGuard}. \textbf{This aligns well with our observation that the metric-based attacks are bounded by the prediction correctness that is further determined by the accuracy gap in Eqn~\ref{pre_correct}, despite the indistinguishable prediction entropy.} Minimizing the training and testing accuracy gap is the most viable approach to further improve the defense effectiveness against these attacks, which further explains why our \textit{NeuGuard} always performs the best among existing defense solutions (see Table~\ref{tbl:metric_attack}).   



\end{document}